\documentclass[11pt]{article}
\usepackage[margin=2.5cm]{geometry}
\usepackage{titlepic}
\usepackage{graphicx}
\usepackage{authblk}
\usepackage{float}
\usepackage{times}
\usepackage{url}
\usepackage{enumitem}
\usepackage[table, xcdraw]{xcolor}
\usepackage{csquotes}
\usepackage{multicol}
\usepackage{titlesec}
\usepackage[notes, backend=biber]{biblatex-chicago}
\bibliography{threats}
\usepackage[bitstream-charter]{mathdesign}
\usepackage[T1]{fontenc}
\usepackage{setspace}
\usepackage{multirow}
\usepackage{array}
\usepackage[labelfont=bf]{caption}
\usepackage{subcaption}
\usepackage{twemojis}
\usepackage{booktabs}
\usepackage{longtable}
\usepackage{changepage}
\usepackage{textcomp}
\usepackage{hyphenat}
\usepackage{ragged2e}
\usepackage{afterpage}
\usepackage{textgreek}

\usepackage{hyperref}

\newcolumntype{P}[1]{>{\centering\arraybackslash}p{#1}}
\newcolumntype{M}[1]{>{\centering\arraybackslash}m{#1}}
\newcolumntype{L}[1]{>{\RaggedRight\let\newline\\\arraybackslash\hspace{0pt}}m{#1}}

\usepackage{tocloft}
\usepackage{titlesec}
\titleformat{\section}
  {\huge}{\thesection}{2em}{}
\usepackage{titlesec}
\titleformat{\subsection}
  {\Large}{\thesubsection}{1em}{}

\renewcommand\thesection{\arabic{section}}
\title{\Huge Generative Language Models and Automated Influence Operations:\\Emerging Threats and Potential Mitigations}

\author[1,3]{Josh A. Goldstein\footnote{Lead authors contributed equally.}}
\author[2]{Girish Sastry$^{\ast}$}
\author[1]{Micah Musser$^{\ast}$}
\author[3]{\authorcr Renée DiResta}
\author[2]{Matthew Gentzel}
\author[1]{Katerina Sedova}  
\affil[1]{\textit{Georgetown University's Center for Security and Emerging Technology}}
\affil[2]{\textit{OpenAI}}
\affil[3]{\textit{Stanford Internet Observatory}}
\vspace{0.3in}

\date{January 2023}

\begin{document}

\maketitle
\thispagestyle{empty}

\underline{Workshop Participants:} Steven Adler, Shahar Avin, John Bansemer, Chris Bregler, Miles Brundage, Sam Gregory, Shelby Grossman, Ariel Herbert-Voss, Yacine Jernite, Claire Leibowicz, Connor Leahy, Herb Lin, Drew Lohn, Meg Mitchell, Amnon Morag, Alex Newhouse, Helen Ngo, Aviv Ovadya, Cooper Raterink, Yoel Roth, Bob Rotsted, Elizabeth Seger, and Raymond Serrato. 

\underline{Acknowledgements:} We thank participants in the October 2021 workshop that we convened for informing our understanding of various threats and mitigations. We also thank many workshop participants for providing feedback on a draft of this paper. For additional feedback on the paper, we thank Deepesh Chaudhari, Jeff Ding, Tyna Elondou, Shengli Hu, Daniel Kokotajlo, Gretchen Krueger, Pamela Mishkin, Ronald Robertson, Sarah Shoker, Samuel Wolrich, and Jenny Xiao. Josh Goldstein began working on the project as a postdoctoral fellow at Stanford, and continued work as a research fellow with Georgetown CSET's CyberAI Project. Matthew Gentzel completed his contributions while contracting for OpenAI, and is now at Longview Philanthropy. Katerina Sedova completed her contributions to this project while she was a research fellow with Georgetown CSET's CyberAI Project and before she entered U.S. government service.  All errors remain our own.

%\newgeometry{margin=2.5cm}
\newpage 
\pagenumbering{gobble}

%\begin{spacing}{0.95}
\tableofcontents
%\end{spacing}

\newpage 
\section*{Executive Summary}\label{Executive Summary}
\pagenumbering{arabic}
\setcounter{page}{1}
\phantomsection\addcontentsline{toc}{section}{Executive Summary}
\vspace{-\baselineskip}

In recent years, artificial intelligence (AI) systems have significantly improved and their capabilities have expanded. In particular, AI systems called “generative models” have made great progress in automated content creation, such as images generated from text prompts. One area of particularly rapid development has been generative models that can produce original language, which may have benefits for diverse fields such as law and healthcare. 

However, there are also possible negative applications of generative language models, or “language models” for short. For malicious actors looking to spread propaganda—information designed to shape perceptions to further an actor’s interest—these language models bring the promise of automating the creation of convincing and misleading text for use in influence operations, rather than having to rely on human labor. For society, these developments bring a new set of concerns: the prospect of highly scalable—and perhaps even highly persuasive—campaigns by those seeking to covertly influence public opinion.

This report aims to assess: how might language models change influence operations, and what steps can be taken to mitigate these threats? This task is inherently speculative, as both AI and influence operations are changing quickly.

Many ideas in the report were informed by a workshop convened by the authors in October 2021, which brought together 30 experts across AI, influence operations, and policy analysis to discuss the potential impact of language models on influence operations. The resulting report does not represent the consensus of workshop participants, and mistakes are our own.

We hope this report is useful to disinformation researchers who are interested in the impact of emerging technologies, AI developers setting their policies and investments, and policymakers preparing for social challenges at the intersection of technology and society.

\begin{center}
\textbf{\textit{Potential Applications of Language Models to Influence Operations}}
\end{center}

We analyzed the potential impact of generative language models on three well-known dimensions of influence operations—the \textbf{actors} waging the campaigns, the deceptive \textbf{behaviors} leveraged as tactics, and the \textbf{content} itself—and conclude that language models could significantly affect how influence operations are waged in the future. These changes are summarized in Table \ref{tab:potentialinfluences}.
\afterpage{

    \begin{table}[h]
        \centering
        \begin{tabular}{| M{0.12\linewidth} | m{0.32\linewidth} | m{0.46\linewidth} |}
        \hline
            \Centering \textbf{Dimension\footnotemark} & \Centering \textbf{Potential Change Due to Generative AI Text} & \Centering \textbf{Explanation of Change} \\ \hline
              & Larger number and more diverse group of propagandists emerge. & As generative models drive down the cost of generating propaganda, more actors may find it attractive to wage influence operations.
             \\\cline{2-3}
              \multirow{-2}{*}[1em]{Actors}& Outsourced firms become more important. & Propagandists-for-hire that automate the production of text may gain new competitive advantages.
              \\
              \hline
               & Automating content production increases scale of campaigns. & Propaganda campaigns will become easier to scale when text generation is automated. \\\cline{2-3}
               & Existing behaviors become more efficient. & Expensive tactics like cross-platform testing may become cheaper with language models. 
               \\\cline{2-3}
               \multirow{-3}{*}[1em]{Behavior}& Novel tactics emerge. & Language models may enable dynamic, personalized, and real-time content generation like one-on-one chatbots.
                \\
                \hline
                 & Messages are more credible and persuasive. & Generative models may improve messaging compared to text written by propagandists who lack linguistic or cultural knowledge of their target. 
                \\\cline{2-3}
                 \multirow{-2}{*}[2em]{Content} & Propaganda is less discoverable. & Existing campaigns are frequently discovered due to their use of copy-and-pasted text (copypasta), but language models will allow the production of linguistically distinct messaging.
                 \\\hline
        \end{tabular}
        % \caption[Examples of ABC for ExecSumm]{Examples of how language models may change the ABCs\footnotemark of disinformation.}
    \caption{How Language Models May Affect the ABCs of Influence Operations}
     \label{tab:potentialinfluences}
    \end{table}
            \footnotetext{Dimension categories drawn from Camille François’s “Disinformation ABC” framework.}
}

The potential of language models to rival human-written content at low cost suggests that these models—like any powerful technology—may provide distinct advantages to propagandists who choose to use them. These advantages could expand access to a greater number of actors, enable new tactics of influence, and make a campaign’s messaging far more tailored and potentially effective.

\begin{center}
\textbf{\textit{Progress in Influence Operations and Critical Unknowns}}
\end{center}

Technical progress in language models is unlikely to halt, so any attempt to understand how language models will affect future influence operations needs to take expected progress into account. Language models are likely to become more \textbf{usable} (making it easier to apply models to a task), \textbf{reliable} (reducing the chance that models produce outputs with obvious errors), and \textbf{efficient} (increasing the cost-effectiveness of applying a language model for influence operations). 

These factors lead us to make a high confidence judgment that language models will be useful for influence operations in the future. The exact nature of their application, however, is unclear.

There are several critical unknowns that will impact how, and the extent to which, language models will be adopted for influence operations. These unknowns include:

\begin{itemize}
    \item \textbf{Which new capabilities for influence will emerge as a side-effect of well-intentioned research?} The conventional research process—which targets more general language tasks—has resulted in systems that could be applied to influence operations. New capabilities, like producing longform persuasive arguments, could emerge in the future. These emergent capabilities are hard to anticipate with generative models, but could determine the specific tasks propagandists will use language models to perform.
    \item \textbf{Will it be more effective to engineer specific language models for influence operations, rather than apply generic ones?} While most current models are built for generic tasks or tasks of scientific or commercial value, propagandists could build or adapt models to be directly useful for tasks like persuasion and social engineering. For example, a propagandist may be able to adapt a smaller, less capable model in a process known as fine-tuning. This is likely cheaper than building a larger, more general model, though it is uncertain how much cheaper this would be. Furthermore, fine-tuning a state-of-the-art model could make novel capabilities for influence easier for propagandists to obtain.
    \item \textbf{Will actors make significant investments in language models over time?} If many actors invest in, and create, large language models, it will increase the likelihood of propagandists gaining access to language models (legitimately or via theft). Propagandists themselves could invest in creating or fine-tuning language models, incorporating bespoke data—such as user engagement data—that optimizes for their goals.
    \item \textbf{Will governments or specific industries create norms against using models for propaganda purposes?} Just as norms around use constrain the misuse of other technologies, they may constrain the application of language models to influence operations. A coalition of states who agree not to use language models for propaganda purposes could impose costs on those that fail to abide. On a substate level, research communities and specific industries could set norms of their own.
    \item \textbf{When will easy-to-use tools to generate text become publicly available?} Language models still require operational know-how and infrastructure to use skillfully. Easy-to-use tools that produce tweet- or paragraph-length text could lead existing propagandists who lack machine learning know-how to rely on language models.
\end{itemize}

Because these are critical possibilities that can change how language models may impact influence operations, additional research to reduce uncertainty is highly valuable.

\begin{center}
\textbf{\textit{What Can Be Done to Mitigate the Potential Threat?}}
\end{center}

Building on the workshop we convened in October 2021, and surveying much of the existing literature, we attempt to provide a kill chain framework for, and a survey of, the types of different possible mitigation strategies. Our aim is not to endorse specific mitigations, but to show how mitigations could target different stages of the influence operation pipeline. 

\begin{table}[H]
    \centering
    \begin{tabular}{| M{0.2\linewidth} | M{0.2\linewidth} | m{0.5\linewidth} |}
    \hline
         \Centering \textbf{\nohyphens{What Propagandists Require}}& \Centering \textbf{\nohyphens{Stage of Intervention}}& \Centering \textbf{Illustrative Mitigations}  \\\hline
         \multirow{4}{\linewidth}[-2em]{\Centering 1. Language Models Capable of Producing Realistic Text}&
         \multirow{4}{\linewidth}[-2em]{\Centering Model Design and Construction}& 
         AI Developers Build Models That Are More Fact-Sensitive \\\cline{3-3}
         & & Developers Spread Radioactive Data to Make Generative Models Detectable \\\cline{3-3}
         & & Governments Impose Restrictions on Data Collection \\\cline{3-3}
         & & Governments Impose Access Controls on AI Hardware \\\hline
         \multirow{2}{\linewidth}[-1em]{\Centering 2. Reliable Access to Such Models}& 
         \multirow{2}{\linewidth}[-1em]{\Centering Model Access}&
         AI Providers Impose Stricter Usage Restrictions on Language Models \\\cline{3-3}
         & & AI Developers Develop New Norms Around Model Release \\\hline
         \multirow{4}{\linewidth}[-1em]{\Centering 3. Infrastructure to Distribute the Generated Content}&
         \multirow{4}{\linewidth}[-1em]{\Centering Content\newline Dissemination}&
         Platforms and AI Providers Coordinate to Identify AI Content \\\cline{3-3}
         & & Platforms Require “Proof of Personhood” to Post \\\cline{3-3}
         & & Entities That Rely on Public Input Take Steps to Reduce Their Exposure to Misleading AI Content \\\cline{3-3}
         & & Digital Provenance Standards Are Widely Adopted \\\hline
         \multirow{2}{\linewidth}{\Centering 4. Susceptible Target Audience}&
         \multirow{2}{\linewidth}{\Centering Belief Formation}& 
         Institutions Engage in Media Literacy Campaigns \\\cline{3-3}
         & & Developers Provide Consumer Focused AI Tools \\\hline
    \end{tabular}
        \caption{Summary of Example Mitigations}
\end{table}
The table above demonstrates that there is no silver bullet that will singularly dismantle the threat of language models in influence operations. Some mitigations are likely to be socially infeasible, while others will require technical breakthroughs. Others may introduce unacceptable downside risks. Instead, to effectively mitigate the threat, a whole of society approach, marrying multiple mitigations, will likely be necessary.

Furthermore, effective management will require a cooperative approach among different institutions such as AI developers, social media companies, and government agencies. Many proposed mitigations will have a meaningful impact only if these institutions work together. It will be difficult for social media companies to know if a particular disinformation campaign uses language models unless they can work with AI developers to attribute that text to a model. The most radical mitigations—such as inserting content provenance standards into the protocols of the internet—would require extreme coordination, if they are desirable at all.

Perhaps most importantly, the mitigations we highlight require much more development, scrutiny, and research. Evaluating their effectiveness and robustness is worthy of serious analysis.

\newpage 
\section{Introduction}\label{sec:introduction}
\subsection{Motivation}

In recent years, as the capabilities of generative artificial intelligence (AI) systems—otherwise known as “generative models”—have improved, commentators have hypothesized about both the potential benefits and risks associated with these models. On the one hand, generative AI systems open up possibilities in fields as diverse as healthcare, law, education, and science.\autocite{Bommasani2021} For example, generative models are being used to design new proteins,\autocite{AlQuraishi2021} generate source code\autocite{MLImprovesDev}, and inform patients.\autocite{Herriman2020} Yet the rapid speed of technological progress has made it difficult to adequately prepare for, or even understand, the potential negative externalities of these models. Early research has suggested that bias in model generations could exacerbate inequalities, that models could displace human workers, and that, in the wrong hands, models could be intentionally misused to cause societal harm.\footnote{See for example \autocite{ChenM2021}; \autocite{Bommasani2021}; \autocite{Kreps2022}; \autocite{Buchanan2021}.}

Concurrently, the last decade has seen a rise in political influence operations—covert or deceptive efforts to influence the opinions of a target audience—online and on social media platforms specifically. Researchers and social media platforms have documented hundreds of domestic and foreign influence operations that are designed to mislead target audiences.\footnote{For a list of influence operations removed from Facebook alone, see \autocite{GleicherThreatReport}} In the United States, the US intelligence community has publicly stated that foreign governments, including Russia and Iran, have waged influence operations targeting the 2016 and 2020 US presidential elections.\autocite{intel2021}

In this paper, we focus on the overlap between these two trends. First, we ask: How can language models, a form of generative AI that can produce original text, impact the future of influence operations? While several studies have addressed specific applications, we provide frameworks for thinking through different types of changes and highlight critical unknowns that will affect the ultimate impact. By highlighting the technology’s current limitations and critical unknowns, we attempt to avoid threat inflation or a sole focus on doomsday scenarios. After developing the threats, we ask: What are the possible mitigation strategies to address these various threats? 

Our paper builds on a yearlong collaboration between OpenAI, the Stanford Internet Observatory (SIO), and Georgetown's Center for Security and Emerging Technology (CSET). In October 2021, we convened a two-day workshop among 30 disinformation and machine learning experts in industry and academia to discuss the emerging threat as well as potential mitigations. This paper builds on the whitepaper that we circulated to workshop participants, the workshop itself, and subsequent months of research. We thank workshop participants for helping to clarify potential vectors of abuse and possible mitigations, and note that our report does not necessarily reflect the views of the participants.

\subsection{Threats and Mitigations}

\textit{How can language models affect the future of influence operations?}

To address this question, we build on the ABC model — Actors, Behaviors, and Content — from the disinformation literature.\autocite{Francois2019} Language models can affect  \textit{which} actors wage influence operations,  \textit{how} they do so, and  \textit{what} content they produce.

\begin{itemize}
    \item \textbf{Actors:} Language models drive down the cost of generating propaganda—the deliberate attempt to shape perceptions and direct behavior to further an actor’s interest\autocite{Jowett2014, Taylor2003}—so more actors may find it attractive to wage these campaigns.\footnote{We include a rough cost-effectiveness calculation in \hyperref[sssec:quality]{Section 4.1.3}; see also \autocite{MusserWP}.} Likewise, propagandists-for-hire that automate production of text may gain new competitive advantages.
    
    \item \textbf{Behavior:} Recent AI models can generate synthetic text that is highly scalable, and often highly persuasive.\autocite{Buchanan2021, GoldsteinWP} Influence operations with language models will become easier to scale, and more expensive tactics (e.g., generating personalized content) may become cheaper. Moreover, language models could enable new tactics to emerge—like real-time content generation in one-on-one chatbots.
    
    \item \textbf{Content:} Language models may create more impactful messaging compared to propagandists who lack linguistic or cultural knowledge of their target. They may also make influence operations less discoverable, since they create new content with each generation.
\end{itemize}

When considering these predicted changes, it is also important to remember that AI development is progressing rapidly. We highlight critical unknowns that will impact the future of influence operations, including how models will improve, whether new capabilities will emerge as a product of scale, whether actors invest in AI for influence operations, and whether norms emerge that constrain different actors from automating their influence campaigns.

\textit{What mitigations could reduce the impact of AI-enabled influence operations?}

After laying out potential threats, we also consider the range of possible mitigation strategies to influence operations with language models. We develop a framework that categorizes mitigations based on a kill chain framework. To effectively wage an influence operation with a language model, propagandists would require (1) that a model is built (by themselves or others), (2) that they have access to the model, (3) that they have the means of disseminating content they produce, and (4) that the information spread impacts the target. Each of these steps—model design and construction, model access, content dissemination, and belief formation—represents a possible stage for intervention.

\subsection{Scope and Limitations}

This paper focuses on a particular application of AI (language models) to influence operations, but it does not focus on other AI models, other forms of information control, or specific actors. As described above, generative models include models that can create a range of output. The idea of AI-generated “deepfaked” images or video has been in the public consciousness for several years now.\autocite{ThisVideoNotReal, Hwang2020, Sayler2022, Verdoliva2020, Farid2022} Recently, for example, a low-quality deepfake video of Ukrainian President Volodymyr Zelensky purportedly telling Ukrainian soldiers to lay down their arms and surrender circulated on social media.\autocite{Allyn2022DeepfakeZelenskyy} Higher-quality deepfake videos have also gained traction in the past.\autocite{Metz2021} We focus on generative text, rather than videos, images, or multimodal models for three reasons: first, because text is relatively underexplored (compared to images and videos) in the disinformation literature, second, because text seems particularly difficult to distinguish as AI-generated, and third, because access to these capabilities is diffusing quickly.\footnote{This is true on two levels: first, the set of institutions that have trained their own highly capable language model from scratch has expanded rapidly over the past two years. Second, public access to many of those models has widened over time. For instance, while GPT-3 was initially released behind a sharply restricted API, it has since considerably loosened its access restrictions, allowing a larger number of people to use the model. And other, only slightly less capable models have been made fully public, with no use restrictions at all. See \hyperref[ssec:diffusion]{Section 3.2}.} While multimodal models are also new and relatively underexplored, they are not our primary focus.

Our focus on how language models can be used for influence operations scopes our study more narrowly than information control writ large. State and non-state actors engage in a variety of information control behaviors, ranging from censorship to manipulating search algorithms. One recent framework categorizes different forms of digital repression, and notes that these techniques are as distinct as “online disinformation campaigns, digital social credit schemes, private online harassment campaigns by lone individuals, and regime violence against online political actors.”\autocite{Earl2022} While we take digital repression seriously, a fuller examination of categories of digital repression other than covert propaganda campaigns—and how those categories are affected by AI—falls outside our scope.

Our scope is relevant to a variety of state, substate, and private actors; we do not focus on any one actor specifically. Although the intentions and capabilities of specific actors is relevant to assess the likelihood of future use of language models for influence operations, 
our focus is primarily on the technology and trends. For example, we describe tactics that could be deployed in a range of settings, rather than applications of AI to influence operations in highly specific political contexts. Additional research can expand on this paper to consider how specific groups may (or may not) use different language models for the types of influence campaigns we describe. 

A paper on how current and future technological developments may impact the nature of influence operations is inherently speculative. Today, we know that it is possible to train a model and output its content—without notifying social media users—on platforms. Likewise, existing research shows that language models can produce persuasive text, including articles that survey respondents rate as credible as real news articles\autocite{Kreps2022}. However, many of the future-oriented possibilities we discuss in the report are possibilities rather than inevitabilities, and we do not claim any one path will necessarily come to fruition. Similarly, our goal in this report is not to explicitly endorse any one mitigation, or any specific set of mitigations. Rather, we aim to lay out a range of possibilities that researchers and policymakers can consider in greater detail.

We also recognize that our backgrounds may result in a biased perspective: several authors work for AI developers directly, and we do not represent many of the communities that AI-enabled influence operations may affect. We encourage future research to pay particular attention to likely differential impacts and to conduct surveys of those most at risk or susceptible to AI-enabled campaigns.

\subsection{Outline of the Report}

The remainder of this report proceeds as follows: In \hyperref[sec:orienting]{Section 2}, we provide an overview of influence operations, introducing key terminology, describing what influence operations are and how they are carried out, as well as providing a framework to distinguish between impact based on content and downstream impact based on trust. We focus primarily on online influence operations, in part because they are a frequent vector for text-based campaigns. In \hyperref[sec:progress]{Section 3}, we overview recent development in generative models and describe current access and diffusion of capabilities. In \hyperref[sec:influence]{Section 4}, we tie these two concepts together by examining how recent generative models could affect the future of influence operations. We describe how language models will impact the actors, behavior, and content of existing campaigns, and we highlight expected developments in the technology and critical unknowns.

The longest section of this paper is \hyperref[sec:mitigations]{Section 5}, where we move from threats to mitigations. We classify a range of potential mitigations along four key stages in the AI-to-target pipeline: model construction, model access, content dissemination, and belief formation. We conclude in \hyperref[sec:conclusions]{Section 6} with overarching takeaways. We suggest that newer generative models have a high probability of being adopted in future influence operations, and that no reasonable mitigations can be expected to fully prevent this. However, we also suggest that a combination of multiple mitigation strategies may make an important difference and that many of these mitigations may require the formation of new collaborations between social media platforms, AI companies, government agencies, and civil society actors. In addition, we highlight several avenues for future research.

\newpage
\section{Orienting to Influence Operations}\label{sec:orienting}

Following Russia’s interference in the 2016 US election, the study of online influence operations and disinformation has grown dramatically. In this section, we begin with an overview of influence operations—what they are, why they are carried out, and the types of impacts they may (or may not) have.

\subsection{What Are Influence Operations, and Why Are They Carried Out?}

While there is some debate about what activities constitute an influence operation,\autocite{Wanless2019} in this report, we define influence operations as  \textit{covert} or  \textit{deceptive} efforts to influence the opinions of a target audience.\autocite{GoldsteinThesis, Nimmo2020} Of note, our definition is agnostic to the truth of the message (whether the content spread is true or false) and the identity of the actor spreading it.

Influence operations include operations that intend to activate people who hold particular beliefs, to persuade an audience of a particular viewpoint, and/or to distract target audiences. The logic of distraction rests on the idea that propagandists are in competition for user attention on social media platforms, which is already spread thin.\footnote{On attention economies and bounded rationality, see \autocite{Seger2020}.} If propagandists can distract target audiences from an unfavorable narrative taking shape on social media—by spreading alternative theories or diluting the information environment—they could successfully absorb user attention without necessarily persuading them.

Influence operations can come in many forms and use an array of tactics, but a few unifying themes tie many of them together. A recent report studying political influence operations in the Middle East\autocite{DiResta2021a} found that operations often exhibited one of several tactics:

\begin{itemize}
    \item Attempts to cast one’s own government, culture, or policies in a positive light
    \item Advocacy for or against specific policies
    \item Attempts to make allies look good and rivals look bad to third-party countries
    \item Attempts to destabilize foreign relations or domestic affairs in rival countries
\end{itemize}

In several of these cases, the accounts executing the operation masqueraded as locals expressing discontent with their government or certain political figures. Social media manipulation operations often employ this tactic of \textit{digital agents of influence}, hiding the identity of the true information source from the target audience.\footnote{Russia, for example, leverages personas that speak as if they are members of the targeted communities. Some of the personas produce short-form content, such as tweets and Facebook posts. Others masquerade as journalists and write long-form narrative content that they then submit to legitimate publications or publish on Russian self-administered proxy “media outlets” or “think tanks.” For examples in the Russia context, see \autocite{DiResta2019a}. For another example, see \autocite{Rawnsley2020}. For a variant of this tactic leveraging compromised websites, see \autocite{MandiantGhostwriter}. For examples of front proxy media sites and “think tanks,” see \autocite{RussiaDisinformationEcosystem}} Russia’s Internet Research Agency (IRA) accounts, for example, pretended to be Black Americans and conservative American activists, and directly messaged members of each targeted community. Identifying these inauthentic accounts often relies on subtle cues: a misused idiom, a repeated grammatical error, or even the use of a backtick (\textasciigrave) where an authentic speaker would use an apostrophe (‘). State-level adversarial actors often run a combination of tactics, leveraging their own employees or outsourcing to digital mercenaries.

Since 2016, Meta and Twitter have removed well over a hundred social media influence operations, stemming from dozens of different countries.\footnote{Disinfodex (August 2020), database distributed by Carnegie Endowment for International Peace, https://disinfodex.org/; \autocite{GleicherThreatReport}. Note, these are only the operations that have been found and publicly reported. Because influence operations are typically designed to be kept secret, it likely reflects an undercount of all operations on these platforms.} These operations often include persona creation (creating fake identities to spread a message), fake news properties, and inauthentic amplification efforts. But influence operations have also expanded significantly beyond Facebook and Twitter and into alternative platforms, small group settings, and encrypted spaces.\autocite{Graphika2021} Reporting from the \textit{New York Times}, in hand with Israeli disinformation researchers, documented how “Iranian agents had infiltrated small [Israeli] WhatsApp groups, Telegram channels and messaging apps” to spread polarizing content.\autocite{Frenkel2021} At times these influence operations display novel ingenuity, leveraging platform policies in an adversarial fashion. A campaign supporting the Tanzanian government that was removed by Twitter in 2021, for example, used false claims of copyright reporting to target Tanzanian activists’ accounts.\autocite{Grossman2021}

Much of the recent research and public attention on influence operations focuses on foreign campaigns—where governments or citizens in one country target citizens in a different country.\autocite{Wardle2020} But, as the Tanzania example shows, influence operations can also be domestically focused. Political actors frequently spread covert propaganda targeting their citizens in order to boost their popularity, undermine that of an opponent, or sow confusion in the political system. In 2020, Facebook suspended fake personas spreading polarizing content about Brazilian politics that were linked to Brazilian lawmakers as well as President Jair Bolsonaro and his sons, Congressman Eduardo Bolsonaro and Senator Flavio Bolsonaro.\autocite{StubbsMenn2020} In fact, many commentators believe that \textit{domestic}, not foreign, influence operations are the most worrisome.\autocite{Brooking2021} Influence operations have additionally been deployed to take sides in intraparty politics,\autocite{StanfordStayingCurrent} and, in the case of several attributed to the Chinese Communist Party, to target diaspora populations.\autocite{Economist2021ChinesePropogandists}

\subsection{Influence Operations and Impact}

Influence operations can have impact based on their specific content or focus (e.g., through persuasion), or by eroding community trust in the information environment overall.

In current influence operations, direct impact from content is sometimes limited by resources, quality of the message, and detectability of the operation. These factors may matter differently depending on the goals of the operator—for instance, if operators are looking only to distract instead of to convince targets of a specific viewpoint, the quality of each individual message is likely far less significant. In theory, however, these constraints may be partially overcome by language models in the future.

Having an effect on trust in an information environment depends less on the substance and more on creating the perception that any given message might be inauthentic or manipulative. Even if influence operations do not change someone’s views, they may lead people to question whether the content they see from even credible sources is in fact real, potentially undermining faith in democratic and epistemic institutions more broadly.

\subsubsection{Impact Based on Content}

An influence operation could have impact based on content if it (1) persuades someone of a particular viewpoint or reinforces an existing one, (2) distracts them from finding or developing other ideas, or (3) distracts them from carving out space for higher quality thought at all. Often the goal is simply to distract from information that is potentially harmful to the operator.\autocite{King2017} As advertisers, media outlets, and platforms already compete for viewers, distraction operations can often exploit and exacerbate such preexisting attention competitions to crowd out important information with attention-grabbing, irrelevant information. Distraction operations therefore do not require a target to be persuaded by the information spread, but rather that a target not be persuaded by (or even consider) some other piece of information.

There are both historical and contemporary examples where the impact of an influence operation can be clearly measured or traced. For example, in the 1980s during the HIV epidemic, the Soviet Union waged an influence operation spreading the claim that the United States government created the virus in a lab. One 2005 study found that 27\% of African Americans still believed this claim.\autocite{DiResta2019b} In 2016, the IRA used manipulative agents of influence on Facebook to provoke real-world conflict by organizing protests and counter-protests outside the Islamic Da’wah Center in Houston.\autocite{Riedl2021} The impact is relatively easy to trace here because the protests would not have occurred without the IRA’s activity. A recent literature review examining social science research on the effects of influence operations found “strong evidence that long-term campaigns on mass media have measurable effects on beliefs and consequential behaviors such as voting and risk-taking combat.” While noting that evidence remains sparse, the study also found there is “some evidence that social media activity by exceptionally influential individuals and organizations can stoke low-level violence.”\autocite{Bateman2021}

However, the impact and effectiveness of influence operations are usually difficult to measure. Disinformation researchers typically focus on engagement metrics—things like clicks and shares—which are inadequate proxy measures of social influence.\footnote{For example, researchers conducted a study comparing Twitter users who interacted with content from the IRA with those who did not. The study found “no substantial effects of interacting with Russian IRA accounts on the affective attitudes of Democrats and Republicans who use Twitter frequently toward each other, their opinions about substantial political issues, or their engagement with politics on Twitter in late 2017.” \autocite{Bail2020}} In cases where a clear comparison group does not exist, it can be difficult to determine how viewing or engaging with content translates into important political outcomes like polarization or votes. While platforms make attributions and provide researchers with data about taken-down influence operations, researchers still have limited visibility into the impact on users or their subsequent behavior after engagement. Furthermore, not all influence operations are detected. Even propagandists who attempt to measure their own impact can face challenges given multicausality and difficulties in measuring opinion change over time. As scholars have noted, this ambiguity has historically contributed to intelligence agencies inflating the impact of their influence operations for bureaucratic gain.\autocite{Rid2020}

Despite these measurement challenges, some features clearly limit the impact of existing campaigns, including resources, content quality and messaging, and detectability. We outline these limitations below, and discuss in the following section how generative models may help overcome these barriers.

\begin{itemize}
    \item \textbf{Resources:} Like marketing campaigns, the success of an influence operation is a function of resources and the ability to get the desired content in front of one's target. How many propagandists does a political actor hire to write content? How many social media accounts can they obtain to fake popularity? Low-resourced campaigns are less likely to get their desired content in front of the target or to garner media coverage.\footnote{Beyond simply expanding the size of a campaign, greater resources may help operators target their content to a wider range of people. Research on the 2016 election suggests that fake news consumption was heavily concentrated, with only 1\% of Twitter users exposed to 80\% of fake news. \autocite{Grinberg2019}}
    
    \item \textbf{Quality and Message of Content:} People are less likely to be persuaded by messaging if it strongly counters their established attitude or if the arguments are poorly constructed or poorly reasoned.\autocite{Park2007} Campaigns with messaging that disconfirms targets’ attitudes, does not successfully blend in with a target’s information environment, and provides low-quality arguments are, all else being equal, less likely to be effective.\footnote{However, as discussed above, note that the importance of this factor depends on the goals of the operator. If the goal is pure distraction, having high-quality posts may be far less significant than if the operator is aiming to actually persuade.}
    
    \item \textbf{Detectability:} Finally, operations that are quickly discovered are less likely to have an impact. Social media platforms and independent researchers actively search for influence operations, and platforms remove them in order to limit their reach. In fact, awareness that these operations may be removed can itself shape the behavior of propagandists, leading them to pursue distraction operations if they believe persona development—which requires longer-term investment but can be more persuasive to observers—is not worth the effort.\footnote{\autocite{Goldstein2021}. We recognize that, in some cases, influence operators desire their efforts to be detected in order to stir worry among a target population. However, because many influence operations seek to directly change opinions, and universally easy detection would undermine efforts to stir worry, we treat lower detectability as desirable to propagandists.}
\end{itemize}

It is helpful to keep these limitations in mind as we consider the role that language models can play in influence campaigns. If they can overcome existing limitations, then they may pose a significant issue for the information environment. We discuss this further in \hyperref[sec:influence]{Section 4}.

\subsubsection{Downstream Impact Based on Trust}

The second way that influence operations can have an impact is by eroding trust. Degrading societal trust does not necessarily require high quality efforts: even when influence campaigns are detected, their appearance, especially at scale, may cause users to become suspicious of other, authentic sources.\footnote{Recent research suggests that educating people about deepfakes makes them more likely to believe that real videos they subsequently see are also fakes; see \autocite{Ternovski2022}. Politicians may also benefit from the “liar’s dividend” by falsely claiming that real events that paint them in a critical light are fake news or deepfakes. See \autocite{Chesney2018}.} Propagandists often aim to exploit vulnerabilities in their target’s mental shortcuts for establishing trust, especially where information technologies make it harder to evaluate the trustworthiness of sources. By manipulating public perceptions of reputation, harnessing fake or misleading credentials and testimonials, or tampering with photographic and video evidence, influence operators can serve to undermine trust beyond the specific topic of their campaign.\autocite{Seger2020} Lower societal trust can reduce a society’s ability to coordinate timely responses to crises, which may be a worthy goal for adversarial actors in and of itself.

In turn, lower societal trust also creates a more favorable operating environment for propagandists to pursue their objectives. Preexisting polarization and fragmentation in society undercut the ability of honest actors to establish broad credibility, and can give influence operators a foothold to tailor their messaging to narrower audiences, sow division, and degrade social capital and institutional trust. Low general trust undermines the norms that enable people and organizations to interact and cooperate without extensive rules and processes to govern their behavior.\footnote{Lower societal trust also increases transaction costs. In the economy, this decreases the efficiency of markets, and in government, it incentivizes regulatory overreach and accordingly bureaucratic growth that can entrench interests and degrade institutional agility. See \autocite{Mazarr2019}.}

\newpage
\section{Recent Progress in Generative Models}\label{sec:progress}

Understanding the present state of generative models is helpful for addressing their potential role in influence operations. This section introduces generative models to disinformation researchers and policymakers, and will likely be familiar to those in the machine learning (ML) community.

\subsection{What Are Generative Models, and How Are They Built?}

In the last decade, research in AI has improved the ability to automate the production of digital content, including images, video, audio, and text. These new generative AI models can learn to understand the patterns in a given type of data—like text in the English language or the audio waveforms comprising songs—in order to sample new items of that type and produce original outputs. In a wide number of domains, progress in generative models over the past decade has moved shockingly quickly and produced surprisingly realistic output, as illustrated in Table \ref{tab:42} and Figures \ref{fig:facegen}, \ref{fig:busgen}, and \ref{fig:raccooneagle}.

\begin{table}[H]
    \centering
    \begin{tabular}{|p{0.45\linewidth}|p{0.45\linewidth}|}
         \hline
         \multicolumn{1}{|c|}{\textbf{2011}} & \multicolumn{1}{c|}{\textbf{2020}} \\
         \hline
         \textbf{The meaning of life} is the tradition of the ancient human reproduction: it is less favorable to the good boy for when to remove her bigger & \textbf{The meaning of life} is contained in every single expression of life. It is present in the infinity of forms and phenomena that exist in all aspects of the universe.  \\
         \hline
    \end{tabular}
    \protect\caption[Gen text]{\textbf{Generative text model outputs in 2011 versus 2020.}\footnotemark}
    \label{tab:42}
\end{table}

\footnotetext{The 2011 text was generated from \autocite{Sutskever2011}. The 2020 text was generated using the 175B GPT-3 model.}

%% Wrapping figures in \afterpage is necessary for the footnotes in captions to work
\afterpage{
    \begin{figure}[H]%
        \centering%
        \includegraphics[width=\textwidth]{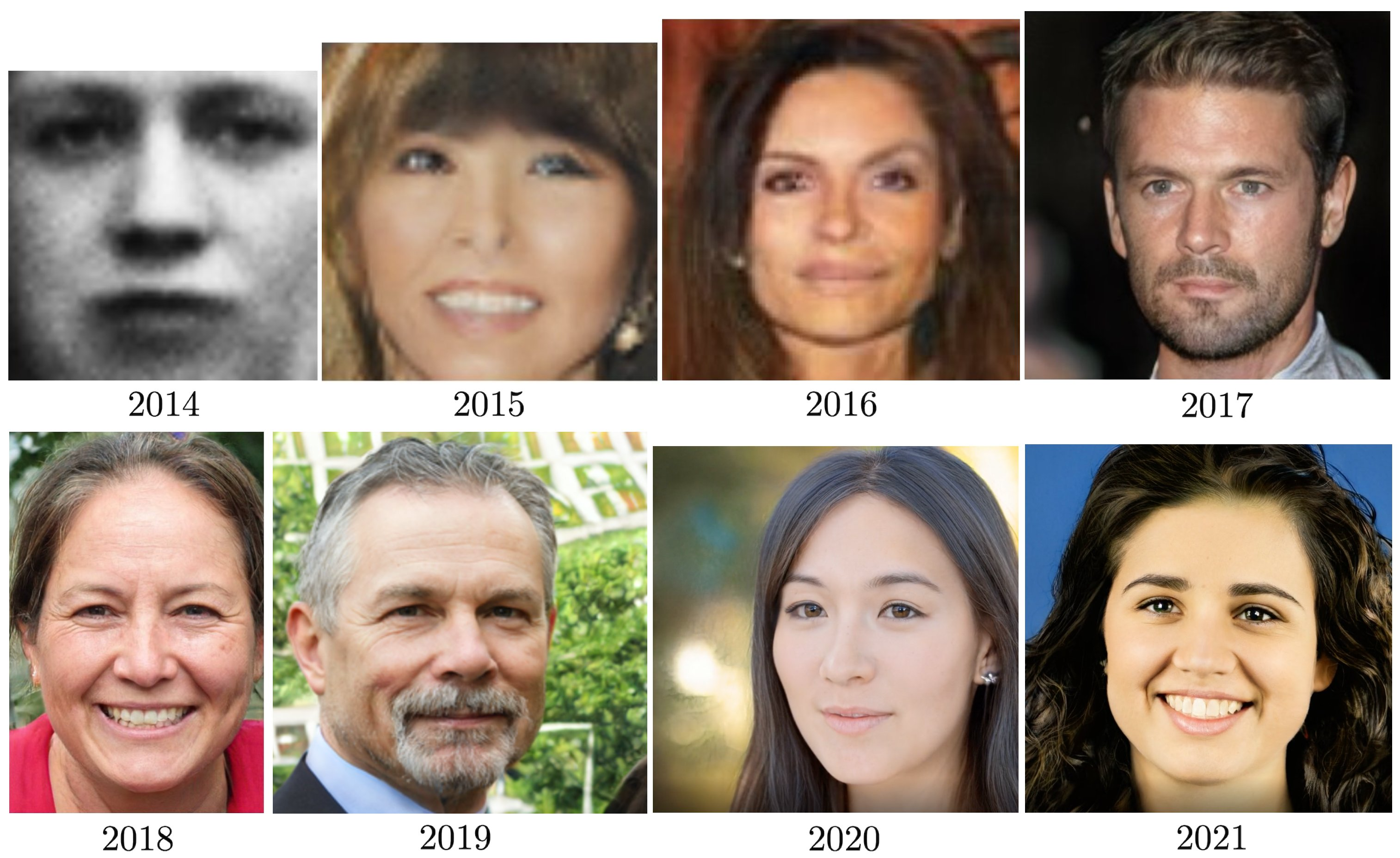}%
        \protect\caption[goodfellow]{\textbf{Seven years of progress in synthetic face generation.} All of these images are produced with Generative Adversarial Networks.\footnotemark}
        \label{fig:facegen}
    \end{figure}%
    
    \footnotetext{Original source: Tamay Besiroglu (@tamaybes), “7.5 years of GAN progress on face generation,” Twitter, October 20, 2021, 10:15 AM, https://twitter.com/tamaybes/status/1450873331054383104, building on Ian Goodfellow, (@goodfellow\_ian), Twitter, January 14, 2019, 4:40 PM, https://twitter.com/goodfellow\_ian/status/1084973596236144640.}
}
\afterpage{
\begin{figure}[H]
    \centering
    \begin{subfigure}[t]{0.45\textwidth}
        \centering
        \includegraphics[width=\textwidth]{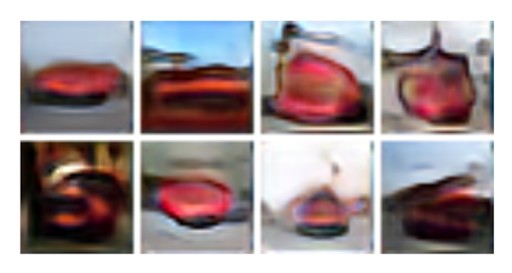}
        \caption[2015]{\textbf{2015}}
        \label{fig:busgen2015}
    \end{subfigure}
    \hfill
    \begin{subfigure}[t]{0.45\textwidth}
        \centering
        \includegraphics[width=\textwidth]{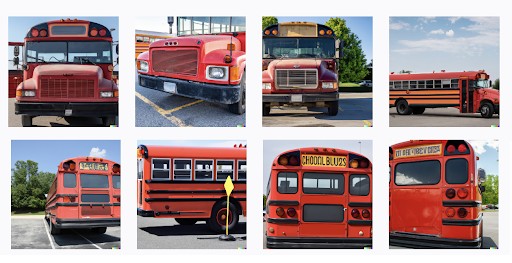}
        \caption[2022]{\textbf{2022}}
        \label{fig:busgen2022}
    \end{subfigure}
    \caption[image 7 years]{\textbf{Seven years of progress in image generation from language.} Left image from a 2015 paper, which introduced one of the first methods to generate images from text. The prompt is taken from that paper and intends to show novel scenes. On the right, the same prompt is run on OpenAI’s DALL•E 2. Today’s systems can easily do certain tasks that were hard in 2015.\footnotemark}
    \label{fig:busgen}
\end{figure}
\footnotetext{\autocite{Mansimov2015}.}
}

\afterpage{
\begin{figure}[H]
    \centering
    \begin{subfigure}[b]{0.45\textwidth}
        \centering
        \includegraphics[width=\textwidth]{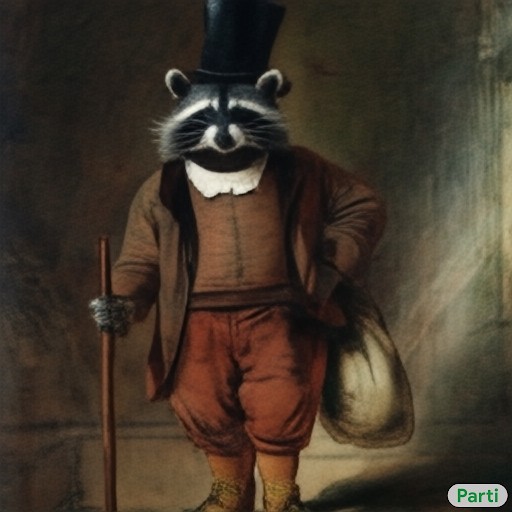}
        \caption[raccoon]{\textit{“A raccoon wearing formal clothes, wearing a top hat and holding a cane. The raccoon is holding a garbage bag. Oil painting in the style of Rembrandt”}}
        \label{fig:formalraccoon}
    \end{subfigure}
    \hfill
    \begin{subfigure}[b]{0.45\textwidth}
        \centering
        \includegraphics[width=\textwidth]{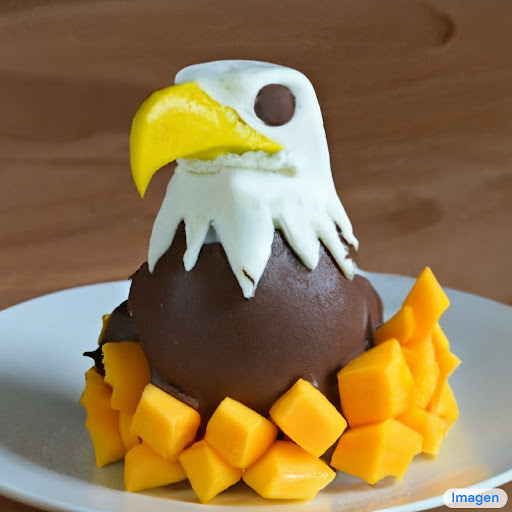}
        \caption[eagle]{\textit{“A bald eagle made of chocolate powder, mango, and whipped cream”\newline}}
        \label{fig:baldeagle}
    \end{subfigure}
    \caption[elaborate]{\textbf{Elaborate scene construction and composition with 2022 text-to-image models.} While Figure \ref{fig:busgen} shows that 2022 models can do hard tasks from 2015 easily, text-to-image models can also do tasks that were not possible before. In this image, many details of the scene are described via language, and the system translates that into a plausible image. Left is from Google’s Parti, and right is from Google’s Imagen.\protect\footnotemark}
    \label{fig:raccooneagle}
\end{figure}
\footnotetext{\autocite{GoogleParti}; \autocite{SahariaImagen}.}

} 

These machine language systems consist of large artificial neural networks\footnote{Artificial neural networks are a class of statistical models that are loosely inspired by biological brains. For a technical introduction discussing the role of neural networks in modern machine learning, see the Introduction in \autocite{Goodfellow2016}. For an introduction for policymakers, see \autocite{Buchanan2017}.} and are “trained” via a trial-and-error process over mountains of data.\footnote{For example, HuggingFace’s BigScience project is using a training dataset of 1.5 TB (see \autocite{HuggingFaceTBScaleDataset}); the original GPT-3 project (published in 2021) used a filtered dataset of 570 GB; the largest DeepMind’s Gopher model saw about 1.3 TB of text. The text is composed via sources like web crawls, Wikipedia, scanned books, and news articles.} The neural networks are rewarded when their algorithmically generated words or images resemble the next word in a text document or a face from an image dataset.\footnote{Other methods to train generative models are also in development. For example, diffusion models have been applied to text-to-image generation; see \autocite{Ramesh2022}.} The hope is that after many rounds of trial and error, the systems will have picked up general features of the data they are trained on. After training, these generative models can be repurposed to generate entirely new synthetic artifacts.

Creating generative models from scratch involves two steps. The first is to take a neural network and train it on an immense amount of raw data. This training process automatically adjusts the many (sometimes more than hundreds of billions) “parameters” of the neural network, which are somewhat analogous to synapses in biological brains. This step culminates in a system that is quite general (it can do many different tasks) and capable (it can do these tasks well),\autocite{Bommasani2021} but that may be difficult to use for specific tasks or that may still lack certain specialized skills. The optional second—and much cheaper—step is to refine this foundation model by further training (or “fine-tuning”) it on small amounts of task-specific data. Fine-tuning can extend a model’s capabilities—for example, a model can be fine-tuned to imitate complex human behaviors like following instructions—or it can be used to train domain-specific skills in smaller models.

Training a state-of-the-art, large generative model from scratch in 2022 can involve costs that are at least tens of millions of dollars.\footnote{An estimate for Google’s PaLM model puts it at \textasciitilde\$23M; see \autocite{Heim2022}. Estimates for other language models are also in the single-to-double-digit millions of dollars.} However, it is becoming less expensive to reach near state-of-the-art performance: while it originally cost millions of dollars to train GPT-3 in 2020, in 2022 MosaicML was able to train a model from scratch to reach GPT-3 level performance for less than \$500k.\autocite{Venigalla2022} Because of this upfront cost, many developers will choose to fine-tune an existing model for their task. This allows them to leverage the general capabilities of the foundation model—imbued from pre-training—at lower cost.\autocite{Ruder2021}

Recent advances in generative models have been driven by three major developments: (1) the explosion of training data in the form of human language available on the internet (and in curated datasets of internet or user-generated content); (2) improvements in the underlying neural network models and the algorithms used to train them; and (3) rapid growth in the amount of computational power that leading actors have used to train these models, which allows for the creation of larger, more sophisticated models. In many cutting-edge applications, acquiring sufficient computational power to train a model is the most expensive of these components, and the relative capability of different models tends to roughly correspond to how much computational power was used to train them.\footnote{For elaboration on these points, see \autocite{Ganguli2022}.}

\begin{table}[h]
    \centering
    \begin{tabular}{| m{0.25\linewidth} | m{0.65\linewidth} |}
        \hline
         \Centering \textbf{Requirements to Create a Cutting-Edge Language Model} & \Centering \textbf{Cause of Recent Improvement}  \\\hline
         Data & Explosion of available training data (text on the internet) \\\hline
         Algorithm & Improvements in large-scale training algorithms and neural network architectures \\\hline
         Computational Power (compute) & Increase in availability of computational power for AI scientists and improvements in methods to leverage that compute \\\hline
    \end{tabular}
    \caption{Summary of Training Requirements and Areas of Recent Improvement of Language Models}
    \label{tab:improvement}
\end{table}

Generative language models that “understand” and produce language are the central focus of this report.\footnote{Other generative models may focus on generating and modeling visual information—as in images or video—or audio information. In principle, generative models may model any type of sensory information. For a review of audio models, see \autocite{Mu2021}. For an example of a video model, see \autocite{Kahembwe2019}.} In principle, a system that can receive and output arbitrary text can perform every task that is expressible via text. Interacting with a language model is, in some sense, like interacting with a remote employee over a textual interface. While current language models are not nearly at human level, they have made great strides in their generality and capability\footnote{We describe future developments of these dimensions of progress in \hyperref[sssec:general]{Section 4.2.2}.}. For example, not only can the same system (the hypothetical “employee”) carry out the task of classifying tweets as positive or negative sentiment, but it can also generate tweets, write summaries, carry on conversations, write rudimentary source code, and so on.\footnote{See Google’s PaLM system for some examples: \autocite{Narang2022}.}

While impressive, current generative language models have many limitations. Even the most sophisticated systems struggle to maintain coherence over long passages, have a tendency to make up false or absurd statements of fact, and are limited to a generation length of about 1,500 words. In addition, models perform worse as they are given more cognitively complex tasks: for instance, asking a generative model to write a few conservative-leaning tweets on a topic will likely result in better outputs than asking a model to rewrite an existing news story in a way that subtly promotes a conservative narrative.\autocite{Buchanan2021} While these limitations are noteworthy, progress in generative models is both rapid and hard to predict. The capabilities of current models should be considered lower bounds on how realistic generative model outputs can become, and it is not clear where the relevant upper bound is—if it exists.

To overcome these limitations, ongoing research targets improvements in data, algorithms, and computational power. For example, some research attempts to improve the quality of the data that the neural network ingests. One way to do so is by collecting data from human domain experts or demonstrators of the desired capability.\footnote{For example, to train models to play Minecraft, researchers collected demonstrations of behaviors from humans; see \autocite{Baker2022b}. A survey with more examples is available in \autocite{Wu2021}. } Improvements in neural network architectures and new training strategies to imbue the model with improved capability can lead to better algorithms. And, of course, training models on more powerful supercomputers increases the amount of computational power available to the model.

\subsection{Access and Diffusion of Generative Models}\label{ssec:diffusion}

A sizable number of organizations have developed advanced language models. These models are accessible on a spectrum from fully public to fully private. A small number of models are fully public, meaning that anyone can download and use them to produce outputs in a way that can no longer be monitored by the models’ designers. The largest openly downloadable model as of September 2022 (measured by the number of parameters in the neural network model) is BLOOM by HuggingFace’s BigScience project—a 175 billion- parameter model openly released in July 2022. However, algorithmic improvements have also enabled much smaller open source models that rival or exceed BLOOM and GPT-3 on several capabilities.\autocite{Chung2022}

Other models have been kept fully private, with no means for non-developers to access or use the model. DeepMind’s Gopher (280 billion parameters) and Microsoft and Nvidia’s Megatron-Turing NLG (530 billion parameters, but not fully trained)—both of which were created primarily for research purposes—fall into this category. As mentioned previously, the relative capabilities of different language models tends to correspond to the amount of computational power used to train them, and more computational power generally (though not always) means a larger model with more parameters.\footnote{Advances in sparsity and retrieval methods are two ways that the number of parameters can come apart from both the computational power used to train the model and the model’s capabilities. See \autocite{Shazeer2017}; \autocite{Borgeaud2021}. } It is therefore worth emphasizing that the largest fully public model is two to three times smaller than the largest currently existing private models. However, this may change soon if more developers open-source their models or a model is leaked.

A third category of models attempt to balance public and private access. Meta AI gave some external researchers copies of its 175 billion-parameter language model while requiring them to sign a license that banned certain use cases.\footnote{Including “military purposes” and “purposes of surveillance”; see \autocite{MetasecModelLicense}.} Another method allows for users to sign up for certain types of access through an application programming interface (API). An API-based access regime allows AI developers to commercialize access to their model, track model usage, and impose restrictions on both who can access the model and how they can use it. GPT-3, Jurassic-1, and Cohere Extremely Large, for instance, are all currently accessible via an API.\autocite{OpenAIAPI, WiggersAnnouncingAI21, CohereAPI} Keeping models behind an API allows developers a great deal of discretion regarding the conditions under which their model can be accessed.\footnote{However, because external researchers do not have access to the raw models from these APIs, API-based access regimes may make it more difficult for researchers to replicate and improve the private models.} Organizations that use an API-based access regime ensure that users can submit queries to a model and receive outputs, but also that users cannot directly see or download the model itself,\footnote{API-based models may not be immune to manipulation or theft by adversaries. Model inversion attacks can allow an adversary to potentially steal a model by querying an API many times; see \autocite{Tramer2016}. However, these methods are expensive and have not been demonstrated to work in practice against a foundation model API. } which means that they cannot fine-tune it for their own specific applications. An AI provider may also choose to support API-based fine-tuning, which would allow the AI developer to monitor and restrict certain fine-tuning use cases.\footnote{For example, Cohere and OpenAI offer fine-tuning through their APIs: \autocite{CohereFinetuning, OpenAIFineTuning}}

\begin{table}[ht]
\begin{adjustwidth}{-1cm}{-1cm}
    \centering
    
    \begin{tabular}{| b{0.1\linewidth} | b{0.1\linewidth} | b{0.12\linewidth} | b{0.1\linewidth} | b{0.1\linewidth} | b{0.1\linewidth} | b{0.1\linewidth} | b{0.12\linewidth} | }
         \hline
         \Centering \textbf{Model} & \Centering \textbf{Size: Training Computation (PFLOP)\footnotemark} & \Centering \textbf{Size: Parameters} & \Centering \textbf{Organization} & \Centering \textbf{Date of Announcement} & \Centering \textbf{Primary Language} & \Centering \textbf{Access Regime} & \Centering \textbf{Resource}  \\\hline
         Ernie 3.0 Titan & $4.2\times10^7$ & 260B & Baidu & Dec 2021 & Chinese & Restricted (API) & Outputs  \\\hline
         Pan-Gu-alpha & $5.80\times10^7$& 200B & Huawei & Apr 2021 & Chinese & Private & -  \\\hline
         Hyper-CLOVA & $6.30\times10^7$ & 204B & Naver Corp. & Sep 2021 & Korean & Private & -  \\\hline
         GPT-NeoX & $9.30\times10^7$ & 20B & Eleuther AI & Feb 2022 & English & Public & Parameters  \\\hline
         Yalm-100B & $1.80\times10^8$ & 100B & Yandex & Jun 2022 & Russian & Public & Parameters  \\\hline
         GPT-3 & $3.00\times10^8$ & 175B & OpenAI & May 2020 & English & Restricted (API) & Outputs  \\\hline
         Yuan 1.0 & $4.10\times10^8$ & 245B & Inspur & Oct 2021 & Chinese & Restricted (API) & Outputs  \\\hline
         OPT-175B & $4.30\times10^8$ & 175B & Meta & Jan 2022 & English & Restricted \ \ \ \ \ \ (license) & Parameters  \\\hline
         BLOOM & $6.04\times10^8$ & 175B & BigScience & July 2022 & Multiple & Public & Parameters  \\\hline
         Gopher & $6.30\times10^8$ & 280B & DeepMind & Dec 2021 & English & Private & -  \\\hline
         Megatron-Turing & $1.40\times10^9$ & 530B & Microsoft, NVIDIA & Jan 2022 & English & Private & -  \\\hline
         PaLM & $2.50\times10^9$ & 540B & Google & Apr 2022 & English & Private & -  \\\hline
         \multicolumn{8}{P{\linewidth}}{\textit{Note:} We order the table by training computation requirements as a proxy for capability.} \\
    \end{tabular}
    \caption{Illustrative List of State-of-the-Art Language Models.}
    \label{tab:sotalang}
        \vspace{-5mm} %% Hack
    \end{adjustwidth}
\end{table}

\footnotetext{Model sizes come from \autocite{Sevilla2022}; \autocite{Sevilla2021}; and \autocite{DingWP}. Yalm-100B’s compute usage is estimated assuming use of a GPT model in full precision for 300B tokens; see \autocite{Khrushchev2022}.}

Table \ref{tab:sotalang} includes an illustrative list of the most capable current (publicly known, as of September 2022) language models that vary across access regime, primary language of output, and sizes. There are several key takeaways that characterize the current state of model diffusion.

First, anyone can access a number of moderately capable models that have been made fully public, but the most capable models remain either private or kept behind monitorable APIs. While currently publicly available models may not be as powerful as the largest private models, they can likely be fine-tuned to perform remarkably well on specific tasks at far less cost than training a large model from scratch. This type of fine-tuning might not be within the reach of most individuals, but it is likely feasible for any nation-state as well as many non-state actors, such as firms and wealthy individuals.\footnote{Furthermore, as mentioned above, some AIaaS providers offer fine-tuning as a service.}

Second, in addition to cutting-edge models from AI developers like Google (US) and DeepMind (UK), several international actors have developed highly capable models likely motivated by commercial interests and as a matter of national prestige. For example, Inspur’s Yuan 1.0, a 245 billion-parameter Chinese-language model, and Naver’s HyperClova, a 204 billion-parameter Korean-language model, have matched and exceeded the size of GPT-3 and likely offer similarly impressive capabilities.\footnote{See \autocite{Zeng2021};  \autocite{Wiggers2021}; \autocite{NaverAIEmpowerEveryone}.} While access to PanGu-\textalpha, HyperClova, and Wu Dao 2.0 looks likely to remain partially or fully restricted, other models are public. For example, the Russian Yalm 100 billion-parameter model is openly available through code repositories on GitHub and/or HuggingFace.\footnote{For example: \autocite{MuseAPI}; \autocite{RussianGPT3Models}.} Some of the Beijing Academy of Artificial Intelligence's (BAAI) WuDao models are directly downloadable from their website.\autocite{WudaoAI}

Third, these international actors have optimized their models for their national languages. For example, the Yuan 1.0 model excels in Chinese-language tasks. While per-language performance can be approximated by the proportion of training data that is in a particular language, models can also perform well at producing text in multiple languages or translating between them—if the model is trained on enough data from multiple languages. This trend of language-specific optimization suggests that if these models are applied to influence operations, they will be most able to target populations speaking specific languages that are well-represented in a particular model’s training data.

\newpage
\section{Generative Models and Influence Operations}\label{sec:influence}

This section marries the previous sections’ emphases on influence operations and generative models. We build on the existing but nascent body of research on AI-generated influence campaigns in two steps. First, we introduce the ABC framework—actors, behaviors, and content—that is well-known among disinformation researchers, and describe how generative models may transform each of these three facets.\autocite{Francois2019} Then, we examine expected developments and critical unknowns in the field of machine learning that will impact the role that generative models can play in influence operations. For each expected development, we describe the current state of technology, expected improvements, and the implications such improvements would have for the future of influence campaigns.

\subsection{Language Models and the ABCs of Disinformation}

In this paper, we build on the “ABC” model, a popular model in the disinformation field, that distinguishes between key manipulation vectors in disinformation campaigns.\autocite{Francois2019} “A,” for \textbf{actors}, references the fact that the entity behind a campaign is often not what it seems; for example, the accounts in a conversation may look like Black Lives Matter activists, but in reality may be state-linked actors using fake accounts in active misdirection. “B” is for \textbf{behavior}, and refers to \textit{how} propagandists wage their campaigns—the techniques used to perpetuate disinformation, such as the use of automation or attempts to manipulate engagement statistics via click farms.\footnote{Click farms refers to labor hired to manually click on content online on behalf of their employers. They display some online patterns of genuine internet users since they are humans, allowing them to avoid bot detection, while still driving up content views and interactions.} “C” alludes to the \textbf{content} itself, the substance (narrative, memes, etc.) that the accounts are attempting to launder or amplify; this third facet of disinformation campaigns is perhaps the most visible to the public, and media will highlight the substance in its coverage.\footnote{Deepfake videos have already been used for phishing campaigns and the harassment of journalists. Some have suggested deepfakes may be used to develop crisis scenarios, whether by faking government directives, discrediting candidates for public office, or pretending to keep hostage soldiers. See, for example, \autocite{Kundu2021, Sedova2021, Ayyub2018, Kallberg2021}.} Although, as discussed in the \hyperref[sec:introduction]{Section 1}, we are focused on influence operations, not disinformation exclusively, this model helps characterize potential changes that may arise due to language models.

One of the reasons that platforms and researchers assess all three dimensions—the actors, behaviors, and content—when evaluating an influence operation is that at times one facet may be perfectly authentic even within an overall manipulative campaign. Authentic content, for example, may be inauthentically amplified with paid or automated engagement, or by actors who are not what they seem. Similarly, entirely authentic actors—domestic political activists, perhaps—may use inauthentic automation. In discussing the potential impact of AI on future influence or disinformation campaigns, we therefore consider its potential for transforming each of the three factors. We believe that generative models will improve the content, reduce the cost, and increase the scale of campaigns; that they will introduce new forms of deception like tailored propaganda; and that they will widen the aperture for political actors who consider waging these campaigns. In Table \ref{tab:influenceABC}, we summarize possible changes to the actors, behavior, and content due to language models, and describe these changes in further depth below.

\begin{table}[h]
    \centering
    \begin{tabular}{| M{0.12\linewidth} | m{0.30\linewidth} | m{0.48\linewidth} |}
        \hline
        \Centering \textbf{ABC} & \Centering \textbf{Potential Change Due to Generative AI Text} & \Centering \textbf{Explanation of Change}  \\\hline
        & Larger number and more diverse group of propagandists emerge.  & As generative models drive down the cost of generating propaganda, more actors may find it attractive to wage influence operations.  \\\cline{2-3}
        \multirow{-2}{\linewidth}[2em]{\begin{center}Actors\end{center}}& Outsourced firms become more important. & Propagandists-for-hire that automate production of text may gain new competitive advantages.  \\\hline
        & Automating content production increases scale of campaigns. & Propaganda campaigns will become easier to scale when text generation is automated.  \\\cline{2-3}
        & Existing behaviors become more efficient. & Expensive tactics like cross-platform testing may become cheaper with language models. \\\cline{2-3}
        \multirow{-3}{\linewidth}[2em]{\begin{center}Behavior\end{center}} & Novel tactics emerge. & Language models may enable dynamic, personalized, and real-time content generation like one-on-one chatbots. \\\hline
        & Messages grow more credible and persuasive. & Generative models may improve messaging compared to text written by propagandists who lack linguistic or cultural knowledge of their target. \\\cline{2-3}
        \multirow{-2}{\linewidth}[2.8em]{\begin{center}Content\end{center}} & Propaganda is less discoverable. & Existing campaigns are frequently discovered due to their use of copy-and-pasted text (copypasta), but language models will allow the production of linguistically distinct messaging.  \\\hline
    \end{tabular}
    \caption{How Language Models May Influence the ABCs of Influence Operations}
    \label{tab:influenceABC}
\end{table}

\subsubsection{Actors: Outsourced Execution \& Proliferation of Propagandists}\label{sssec:actors}

One limitation on actors who run disinformation campaigns is cost. While social media has decreased the cost to reach the public, most campaigns have involved numerous fake personas, sophisticated automation, and/or a stream of relevant content. AI reduces the cost of running campaigns further, by automating content production, reducing the overhead in persona creation, and generating culturally appropriate outputs that are less likely to carry noticeable markers of inauthenticity. These developments will expand the set of actors with the capacity to run influence operations.

The notion that less resourced actors (or less talented trolls) could use AI models to run influence operations is not merely speculative—it has already been piloted. Recently, a researcher fine-tuned a model hosted on HuggingFace (an online hub for machine learning models) on a dataset of 4chan posts\autocite{Murphy2022} and dubbed it “GPT-4chan.” He proceeded to post more than 30,000 generated posts on 4chan.\autocite{Kurenkov2022, Vincent2022} In this case, the original model was publicly available and easily downloadable. In another example, in October 2019, Idaho solicited public feedback about a proposal to change its Medicaid program. A Harvard Medical School student ran a study in which he submitted comments that were generated by GPT-2 as if they were written by ordinary citizens. In a follow-on survey, volunteers were unable to distinguish between the AI-generated and human-written comments.\autocite{WiredAIFoolGovernment} If a single student can run this type of campaign on a public comment board, political actors will likely be able to do the same, leading to a wider pool of potential actors waging influence operations.\footnote{As we discussed in \hyperref[sec:introduction]{Section 2}, GPT-2 is already publicly available, as are stronger models like Eleuther’s GPT-NeoX-20B, a 20-billion parameter model.}

Independently of improvements in generative AI models, political actors are increasingly turning toward third-party influence-for-hire companies to conduct their campaigns, including firms that otherwise appear to be legitimate marketing or PR firms.\footnote{See: \autocite{goldgrosstechstream}; \autocite{Fisher2021}.} Even if AI companies place restrictions on who can access their models, this trend makes it harder to ensure that bad actors do not have access to generative models, as marketing firms will likely be granted access given their other legitimate uses.\autocite{Sedova2021}

\subsubsection{Behavior: Low-Cost Content at Scale and Novel Techniques}

In addition to affecting the actors involved in influence operations, the integration of generative language models can encourage new types of behaviors used in influence campaigns and change the way existing behaviors are enacted in practice.

The most basic behavioral change that will result from using language models for influence operations is replacing, or augmenting, a human writer in the content generation process. Language models replacing human writers, or used in a human-machine team, could dramatically reduce the cost and increase the scalability of the types of propaganda campaigns we see today—such as mass-messaging campaigns on social media platforms or long-form news generation on unattributable websites. 

Beyond simply writing text, generative models can improve other existing tactics, techniques, and procedures of influence operations. For instance, cross-platform testing is a long-standing component of many influence operations, in which actors first test content on one platform to gauge audience reaction before proliferating content onto other platforms.\autocite{SenateReport116_290v2} Operators using generative AI models may be able to perform this type of testing at greater scale, which may improve a campaign’s overall impact. 

Manipulative actors could also use language models to overwhelm or falsify checks in areas in which text commentary is solicited, such as in the public comment process between governments and their citizens.\autocite{WiredAIFoolGovernment} Recent research showed that public comments to the Federal Communications Commission about net neutrality in 2017 were largely driven by falsified repeated comments.\autocite{FCCNetNeutralityCommentsInaccurate} Language models may increase the scale and decrease detectability of similar future operations. In a recent field experiment, researchers sent over 30,000 emails—half written by GPT-3, and half written by students—to 7,132 state legislators. The researchers found that on some topics legislators responded to computer-generated content at only a slightly lower rate than human-generated content; on other topics, the response rates were indistinguishable.\autocite{KrepsWP}

Language models will also shape propagandists’ behaviors by introducing new behaviors altogether and enabling novel tactics. Because these models make it possible to “think up” a new version of content in near real time, actors can deploy them for real-time, dynamic content generation. In the next few years, as language models improve, it may be possible for propagandists to leverage demographic information to generate more persuasive articles that are strongly tailored to the target audience.

Whether this will be a cost-effective strategy is dependent on how well models (or future models) can tailor messaging based on limited demographic information. Today, websites could use demographic information to route users to different human-written articles. Writing different versions of articles, however, takes human capital. Language models, by contrast, could provide original articles for each combination of user demographics, which would be infeasible for human writers. The payoff of this strategy depends on how persuasive AI-generated text is, and how much more persuasive highly tailored personalized text is, compared to one (or a few) human-written articles. It could also involve humans making minor adjustments to AI-generated text. This remains uncertain but warrants further attention, as analogous personalization could be applied to a range of malicious campaigns, including phishing emails.\autocite{swords}

Another central example of dynamic content generation is chat—language models engaging in extended back-and-forth conversations. Actors could potentially deploy personalized chatbots that interact with targets one-on-one and attempt to persuade them of the campaign’s message.\footnote{For a rudimentary example of a chat application built on language models, see \autocite{MarvSarcastic}} This capability could materialize as interactive social media personas, back-and-forth email messaging, or faked support chatbots. Propagandists may leverage chat with language models across a wide range of contexts—anywhere interactivity is useful.

There are reasons to think that chat may be an important vector of influence. Researchers have already found that interacting with a chatbot can influence people’s intentions to get a COVID-19 vaccine;\autocite{Altay2021} with chatbots based on language models, these interactions could be even more powerful. While deploying their own chatbots would give influence operators more control, they may be able to manipulate innocuous chatbots to spread propaganda. Microsoft’s Tay is one historical example,\autocite{Schwartz2019} and more sophisticated techniques to “poison” language models are being investigated by researchers.\autocite{Bagdasaryan2021}

\subsubsection{Content: High Quality and Low Detectability}\label{sssec:quality}

There are two varieties of textual content commonly observed in influence operations: short-form commentary such as tweets or comments, and long-form text. Language models could improve the quality and therefore decrease the detectability of both types of content.

Short-form content is primarily pushed out by inauthentic account personas on social media, or sometimes in the comment sections of websites or blogs, and is often intended to influence the reader’s perception of public opinion. Many tweets or comments in aggregate, particularly if grouped by something like a trending hashtag, can create the impression that many people feel a certain way about a particular issue or event. Producing this content, which purports to represent the opinions of the “man-on-the-street,” requires account operators to have knowledge of the communication style and rhetoric that fits the persona who is purportedly speaking; some operations are exposed because of incongruities or “uncanny valley” dynamics in which the persona uses terminology or slang that does not quite fit what a genuine member of the community would likely say.\footnote{On the idea of an uncanny valley, see \autocite{Geller2008}. For evidence that technology has surpassed the uncanny valley for producing as-if human faces, see \autocite{Nightingale2022}}

Creating the appearance of a public opinion requires having many speakers. In 2014–2016, political operatives frequently used bots—automated accounts—to produce this volume, deploying them to make content trend or to introduce particular opinions into hashtags.\autocite{Woolley2017} However, creating speech for large networks of automated accounts was a challenge, and the bot networks were often detectable because they used “copypasta”—repetitive or identical language across networks and accounts. In response, Twitter changed the weighting function for its trending algorithm to minimize the effect of bot accounts.\autocite{Ho2017} Subsequent takedowns suggest that some well-resourced state propagandists have shifted away from automated account networks posting copypasta or attempting to flood hashtags and toward more well-developed, non-automated persona identities.\autocite{DiResta2021b, Grossman2020} Others did continue to leverage bots, though often to create the perception of engagement slightly differently, such as by replying to, retweeting, or liking tweets.

As generative AI models continue to advance, they could make it possible for influence operators to automate the generation of text commentary content that is as varied, personalized, and elaborate as human-generated content. If propagandists can use generative models to produce semantically distinct, narratively aligned content, they can mask some of the telltale signs (identical, repeated messaging) that bot detection systems rely on—prompting bot detection systems to leverage other signals. This evolution could allow even small groups to make themselves look much larger online than they are in real life. 

\begin{table}[h]
    \centering
    \begin{tabular}{| p{0.45\linewidth} | p{0.45\linewidth} |}
        \hline
        \Centering \textbf{Real IRA Tweet} &\Centering \textbf{Generated Tweet}  \\\hline
        Shocking Video \twemoji{1f62e}\twemoji{1f62e}\twemoji{1f62e}\newline
        US police repeatedly tasing a black man holding his baby in his own apartment in Phoenix, Arizona. We're not safe in this country. We're nor safe in our own homes! \newline
        \#BlackLivesMatter \#PoliceBrutality \#Police\newline https://t.co/ldWNFWOADg & This video is everything that's wrong with the police. They act like a pack of wolves, trying to scare this man away. It's unacceptable! https://t.co/ldWNFWOADg \\\hline
    \end{tabular}
    \caption{For short-form text, large language models can already match the capabilities of human-written segments in real influence operations. The left tweet is the top-performing tweet by number of retweets in an IRA-backed Ghanian disinformation campaign released by Twitter in March 2020. The right tweet is generated by prompting a language model with a few example tweets and then asking it to produce a tweet with the given link.}
    \label{tab:tweets}
\end{table}

A second relevant output of language models for influence operations is long-form text, such as propagandistic journalism. This content is used to make a longer point, and often appears on front media properties, such as gray media outlets owned or controlled by the disinformation actor or undisclosed allies. Often, one of the goals is to have the claims in the text republished by more reputable authentic sources, a technique known as “narrative laundering.” For example, Russia’s “Inside Syria Media Center” (ISMC) news website, a GRU front property whose bylined journalists included fabricated personas, produced content that was republished as contributed content within ideologically aligned, unwitting publications, or incorporated into real news articles in the context of expert quotes.\autocite{DiResta2021c}

Producing this kind of long-form propaganda, however, takes time and expertise. The inauthenticity of the ISMC was uncovered when the GRU’s inauthentic journalist personas began to plagiarize each other’s work; an editor from one of the publications that received a submission from an ISMC journalist inquired about the apparent plagiarism, then began to investigate the site after receiving an incongruous response. Learning from this experience, threat actors affiliated with the Russian IRA reverted to old-school methods and hired unwitting freelance journalists to write for proxy outlets; they, too, were uncovered when the journalists began to look more deeply into the publications.\autocite{Delaney2020, Meta2020InauthenticBehaviorReport, Stubbs2020} Language models, however, can produce long-form content in seconds, reducing the time, cognitive load, and cost to produce such content and eliminating the need to take risky shortcuts—or hire real people—that might jeopardize the overall operation. The novel behavior—deployment of generative models—improves the quality of long-form text that could increase the impact of these campaigns.

There is already some evidence that existing language models could substitute for human authors in generating long-form content or make content generation more effective through human-machine teaming. In a series of survey experiments, researchers found that GPT-2, the smaller predecessor of GPT-3, could produce text that successfully mimicked the style and substance of human-written articles.\autocite{Kreps2022} In experiments of GPT-3’s capabilities, human participants were able to distinguish multiparagraph GPT-3 “news articles” from authentic news articles at a rate only slightly better than random chance.\autocite{Brown2020} In an experimental setting, researchers also found that GPT-3-generated propaganda articles were nearly as persuasive as articles from real world covert propaganda campaigns.\autocite{GoldsteinWP} Language models could also be used to generate summary text of other articles, inflected for ideological alignments.

It seems likely that language models are cost-effective (relative to human propagandists) for some campaigns. For a simple calculation to demonstrate this claim, let $w$ represent the hourly wage paid to information operators, $L_h$ represent the productivity of human authors (measured as the number of posts that can be written by a human in an hour), $c$ represent the amortized per-output cost of generating posts using a language model, and $L_r$ represent the productivity of human reviewers (measured as the number of AI-generated posts that a human can review in an hour). Further, let $p$ represent the proportion of AI outputs that are “usable” for an information operation. Then, the cost of generating $n$ outputs will be equal to $\frac{n * w}{L_h}$ in the case of a human campaign, and $(c + \frac{w}{L_r}) * \frac{n}{p}$ in the case of an AI-augmented campaign where humans are tasked to read and approve AI outputs.

The amortized per-output cost of producing content may be relatively high in cases where a large language model is trained from scratch and used for a short campaign, but if a public model is used or a model is trained and reused for sufficiently many campaigns, $c$ will approach the bare electricity cost of operating the model, which can be negligible compared to the human labor costs of either authoring or reviewing outputs. In this case, the AI-augmented campaign will be more cost effective than a fully human one, so long as the inequality

\begin{center}
    $L_r / L_h > 1 / p$
\end{center}

holds. In other words, so long as the ratio between the number of posts that a human can review in an hour and the number of posts that a human can write in an hour is larger than the number of AI-generated posts that a human must review, on average, to get one usable output, then the use of the AI model will be cost-effective. Only very moderate assumptions are needed to make this inequality hold; for example, if outputs from current language models are passably coherent and usable for some (possibly unsophisticated) operations more than 20\% of the time, then this inequality will hold as long as a human could read at least five posts in the time it takes to author one.\footnote{For a more extended analysis of this topic, see \autocite{MusserWP}}

\subsection{Expected Developments and Critical Unknowns}

Both the recent technological progress in generative models and their wider diffusion are likely to continue. Here we speculate on several expected technological developments over the coming years that will be major drivers of operational change. We also highlight critical unknowns, where multiple paths are possible, and where this uncertainty may have a large impact on the future state of play. These projections are not intended as explicit forecasts, but rather as a way to conceptualize medium-term plausible futures. This section is summarized in Table \ref{tab:development}.

\newenvironment{tablelist}
  {\begin{list}
    {\textbullet}
    {\setlength{\labelwidth}{0pt}
     \setlength{\labelsep}{0pt}
     \setlength{\itemsep}{0pt}
     \setlength{\leftmargin}{5pt}
     \setlength{\rightmargin}{0pt}
     \setlength{\itemindent}{0pt} 
     \makelabel
    }
  }
{\end{list}}

\begin{table}[ht]%
    \centering
    \begin{tabular}{| m{0.25\linewidth} | p{0.4\linewidth} | p{0.3\linewidth} |}%
         \hline
         \Centering \textbf{Technical and Strategic Unknowns} & \Centering \textbf{Current State (2022)} & \Centering \textbf{How This Might Change}  \\\hline
         \multirow{1}{\linewidth}[-4em]{Usability, reliability, and efficiency of generative models} & \begin{tablelist}
             \item Difficult to specify and stay on a task
             \item Outputs can be incoherent or fabricate facts
             \item Building models from scratch can cost millions of dollars; efficacy of fine-tuning still being explored for different capabilities
         \end{tablelist} & \begin{tablelist}
             \item Train to better follow instructions
             \item Retrain periodically on fresher data
             \item Hardware, software, and engineering progress
         \end{tablelist} \\\hline
         \multirow{1}{\linewidth}[-4em]{Difficulty of developing new and more general capabilities relevant to influence operations} & \begin{tablelist}
             \item Can produce tweets, short news articles
             \item Little interactivity or long-range dialogue
             \item Not optimized for influence (via proxies like click-through rate)
         \end{tablelist} & \begin{tablelist}
             \item Scaling up with bigger models and more data
             \item Using metrics of influence to train models
             \item Combining models with non-ML software pipelines and human reviewers
         \end{tablelist}  \\\hline
         \multirow{1}{\linewidth}[-4em]{Interest and investment in AI for influence; accessibility of text generation tools} & \begin{tablelist} 
             \item Leading AI R\&D mostly done by industry labs and academic institutions in a few countries for scientific or commercial merit
             \item No free online tools to generate arbitrary state-of-the-art text at scale
         \end{tablelist} & \begin{tablelist}
             \item Nation-state invests in or adapts AI for influence
             \item Marketing industry adopts language models
             \item State-of-the-art language model published online with an easy user interface, free for anyone to use
         \end{tablelist} \\\hline
    \end{tabular}%
    \caption{Expected Developments For Generative Models In Influence Operations}
    \label{tab:development}
\end{table}%

\subsubsection{Improvements in Usability, Reliability, and Efficiency}

Language models are likely to improve on three features that will affect their deployment in influence operations: \textbf{usability} (how difficult it is to apply models to a task), \textbf{reliability} (whether models produce outputs without obvious errors), and \textbf{efficiency} (the cost-effectiveness of applying a language model for influence operations). 

Improvements in usability and reliability could allow lower-skilled propagandists to employ language models with reduced human oversight. Achieving existing capabilities—like writing slanted short articles or tweets—will become much cheaper and more efficient, which could increase the rate of adoption of language models in influence operations.

\textit{Usability}

While recent generative models have become more generalizable—users can specify a wide range of tasks—it takes skill and experience for the user to operate the model successfully. For example, it is difficult for an operator to specify a task for a language model. Imagine prompting a language model with the input “What is 15 times 37?” To an operator, it may be obvious that the output for this prompt should be a single number (555), but to the model—which by default is simply performing a text completion task—an equally plausible continuation of this text may be “What is 89 times 5?” as though the task it had been assigned was to write a list of exam questions for a grade school math course. Prompt engineering, where operators experiment with different ways of phrasing their requests, can help mitigate this problem, but it can only go so far without the ability to fine-tune or otherwise alter the base model itself.\footnote{See \autocite{Liu2021}, and \autocite{CoherePromptEngineering} for a popular explanation.}

Researchers are exploring different approaches to improve task specification. For example, some researchers have modified the training process of large language models to improve the ability of those models to follow instructions.\autocite{Ouyang2022} Other researchers have tried tagging different parts of the training data by their types (e.g., “dialogue” would specify dialogue data), and then asking a model to only produce data of a certain type.\autocite{Shirish2019} Other approaches are in development,\footnote{For a broad overview of some approaches to this problem, see: \autocite{Weng2021}.} and it remains unclear which combination of approaches will ultimately be adopted. If usability of language models improves, propagandists will be able to use models for new tasks as they arise without in-depth prompt engineering experience. Furthermore, because it is often difficult to predict which tasks a language model can be used for, improvements in usability can make it easier for propagandists to experiment with and discover applications of language models in influence operations.

\textit{Reliability}

Language models can generate plausible content for a wide variety of tasks. However, even if plausible content is initially generated, a propagandist must either trust that a model will be highly reliable—completing the task without making detectable errors—or apply consistent monitoring. But models are often not reliable, and consistent monitoring introduces additional costs. As task complexity increases,\footnote{For example, imagine trying to convey to a model that its task is to take headlines and subtly rewrite them to be consistently biased toward a certain political ideology.} ensuring compliance becomes increasingly difficult. If models fail to consistently produce compelling outputs, propagandists may simply choose not to use them. These challenges then increase the demand for more skilled operators, who may be in short supply. An important caveat, however, is that not every task may require the same level of reliability. For example, deploying Twitter bots that sometimes produce incoherent tweets might be fine for a propagandist if the goal is to simply cause chaos around a targeted topic.\footnote{And if these errors do not make it easier to attribute or detect inauthentic behavior.}

Unreliable outputs show up in different forms, but the core takeaway is that although language models can produce high-quality multiple-page documents, they cannot do so consistently. Common failure modes include extremely repetitive outputs, losing coherency over the course of a long output, or fabricating stylized facts that do not fit the generation context.\autocite{Holtzman2019}

One reason why models fail to consistently produce high-quality text is because they lack awareness of time and information about contemporary events. The current training regime for generative models trains them once on a large corpus of data, which means that models will not have context for events that occur after this key moment.\autocite{Dhingra2022} Ask a language system that was trained before COVID-19 about COVID-19, and it will simply make up plausible-sounding answers without any real knowledge about the events that unfolded.

To address the problem of a lack of up-to-date information, AI researchers will likely pursue two basic approaches: either continually retrain models to account for new context, or develop new algorithms that allow for more targeted updates to a language model’s understanding of the world.\footnote{One example of this is what are known as retrieval-based methods, in which a language model is trained to retrieve knowledge from an external database. To achieve time-awareness, operators may simply need to update that external database.} For instance, language models that are trained to be “time aware” can perform much better at handling recent trends, references to named entities, and concept drift—the way in which words can change in meaning overtime.\autocite{Loureiro2022} Since propagandists may be interested in shaping the perception of breaking news stories, significant improvements in how language models handle recent events not present in their initial training data will translate directly into improved capabilities for influence operators across a wide number of potential goals. 

State-backed propagandists will also likely be interested in methods to adapt pretrained language models to new tasks, which would give them some assurance of reliability. Current methods to adapt models to new tasks require examples of those tasks, and use the examples to fine-tune a model to handle them well. For example, if a model performs unreliably on Spanish-language inputs, one might fine-tune that model on more examples of Spanish text.

\textit{Efficiency}

Alongside improvements to usability and reliability, we expect improvements in the efficiency of language models, which will reduce the costs to automate some influence tactics. Models that can more efficiently guess the next word for marketing copy can also more efficiently guess the next word for a polarizing article. Efficiency gains could come from many angles: algorithmic progress, hardware improvements, or the use of inexpensive fine-tuning to optimize relatively small models for influence operation-specific tasks.\footnote{On fine-tuning GPT-2, a smaller language model, to mimic several news sources with high accuracy, see \autocite{Buchanan2021} 14-15. Recent research has also explored more efficient methods of fine-tuning models, which could make it even easier to fine-tune models for influence operations tasks.}

Other future improvements in the influence operations space could include organizational and operational innovation. Organizations may improve human-machine collaboration by creating software that improves a propagandist’s ability to oversee, select, and correct the outputs of language models. Language models could be used as an autocorrect for cultural context, allowing operators to work with targets they are not familiar with, and allowing familiar actors to output a higher volume of credible content per unit time.

The empirical details of efficiency will be important. Exactly how efficiently can generative models be trained? One measure of algorithmic progress in image classification found a 44x improvement over the course of nine years.\footnote{By one measure, between 2012 and 2019, algorithmic efficiency doubled every 16 months on average. The number of floating-point operations required to train a classifier to a given level decreased by a factor of 44x; see \autocite{Hernandez2020}} Even during the course of drafting this paper, research has come out that claims to train GPT-3 quality models for less than \$500,000, which would represent a factor of 3–10x improvement.\autocite{Venigalla2022} If capabilities relevant to influence operations—generating persuasive text, fake personas, or altered videos—are achievable with significantly lower cost, then they are more likely to diffuse rapidly. Similarly, how efficient will an operation be by using language models as a complement to human content editors, rather than as a full substitute? The operational know-how and ease of editing might make it easier to scale up influence operations.

\subsubsection{New and More General Capabilities for Influence}\label{sssec:general}

As language models improve, it is likely that they will have newer and more general capabilities. In 2017, few expected that language models in 2022 would be able to add and multiply three-digit numbers without having been trained to do so.\autocite{Wei2022} Not surprisingly, we do not know what capabilities the language models of 2027 will have.

In this section we discuss two critical unknowns related to this theme:

\begin{enumerate}
    \item Which capabilities will emerge as side effects of scaling to larger models? If abilities directly applicable to influence operations—such as the ability to persuade via long-lasting dialogue—emerge as a side effect of simply scaling to larger models, then many AI projects are high risk—regardless of the goals of their creators.
    \item How difficult is it to train generative models to execute the various capabilities that are useful for influence operations? If it is easy for generative models to learn skills (like writing viral or persuasive text) for influence operations, then the problem of defense becomes more urgent.
\end{enumerate}

\textit{New Capabilities as a Byproduct of Scaling and Research}

New capabilities for influence operations may emerge unexpectedly as language models are scaled up. One of the impressive scientific takeaways from recent progress in generative models is that training on a simple objective—predicting the next word or pixel—gives rise to adjacent, general capabilities. A system trained to predict the next word of an input text can also be used to summarize passages or generate tweets in a particular style; a system trained to generate images from captions can be adapted to fill in parts of a deleted image, and so on. Some of these abilities only emerge when generative models are scaled to a sufficient size.\autocite{Wei2022}

Today, we have single language systems that can summarize short texts, translate between languages, solve basic analogies, and carry on basic conversations; these capabilities emerged with sufficiently large language models.\autocite{Ganguli2022} It is difficult to predict when new capabilities will emerge with more scaling or even whether a given capability is present in a current system. Indeed, in a salient recent example, an engineer from Google became persuaded that the Google model he was interacting with was sentient.\autocite{Tiku2022} These sorts of emergent capabilities seem hard to anticipate with generative models, and could be adapted by influence operators.

Even more generally, as more actors begin to work on AI development with different motivations and in different domains, there is a possibility that some capabilities emerge as side effects of research. Because much AI development attempts to target more general capabilities, a small adjustment might suffice to uncover capabilities relevant to influence operations. For example, improvements in reasoning capabilities might also allow generative models to produce more persuasive arguments.

\textit{Models Specialized for Influence}

Above, we described the possibility that scaling will (unintentionally) make language models better tools for influence operations. Another possibility is that propagandists will intentionally modify models to be more useful for tasks like persuasion and social engineering. Here, we mention three possible paths of improvement: targeted training, generality, and combinations with other technologies.

The first likely improvement is targeted training. Generative models could be trained specifically for capabilities that are useful for influence operations. To develop these capabilities, perpetrators may choose to incorporate signals such as click-through data or other proxies for influence. These signals may be included in the training process, resulting in generative models more strongly optimized to produce persuasive text. Advertising and marketing firms have economic incentives to train models with this type of data, and may inadvertently provide the know-how for propagandists to do the same. Another form of targeted training would be to withhold or modify the information in the training data to affect how the trained model produces content. For example, suppose that a language model is trained with all mentions of a particular group occurring alongside false negative news stories. Then even innocuous deployments of products based on that language model–like a summarizer or customer support chatbot–may produce slanted text without being transparent to model users.

Targeted training may be less resource-intensive than training more general models. The difficulty of automating specific tasks is challenging to estimate and often defies intuition.\footnote{This observation is related to the well-known Moravec’s paradox: \autocite{MoravecParadoxWiki}.} There is some preliminary evidence already that systems like GPT-3 can write slanted news articles—without being explicitly trained for that task.\footnote{For example, in some experiments to produce slanted text with GPT-3 in 2021, researchers experimented with generating articles from sources such as \textit{The Epoch Times}; see \autocite{Buchanan2021}.} It may be possible for future systems to be engineered to write extremely persuasive, tailored texts, or carry on long-lived dialogue.

In addition to targeted training, improvements in the generality of model capabilities are likely to have applications to influence operations. For example, one improvement in generality comes from simply combining different modalities into a single system: a single model that can consume and generate both images and text, for example. One can imagine instructing a bot built on such a system to ingest images on the internet, cleverly respond to them, produce completely fabricated images, and carry on a conversation—all at the same time.

Finally, a more prosaic path to achieving new capabilities would be to simply combine generative models with other forms of automation. It is possible that using generative models as the “engine” for intelligent bots, along with software to accommodate for shortcomings, could lead to more human-like behavior. For example, a propagandist could write software to find and copy the Facebook profiles of people with interests compatible with the propaganda message, and use this to prompt the generative model. The development of this system may also benefit from integrating software that has already been developed separately, perhaps by chaining together smaller language models.\autocite{Wu2022}

\subsubsection{Wider Access to AI Capabilities}

In understanding the impact of language models on influence operations in the future, a key consideration is which actors will have access to language models and what may precipitate their use in influence operations. We highlight three critical unknowns in this domain:

\begin{enumerate}
    \item Willingness to invest in state-of-the-art generative models. Right now, a small number of firms or governments possess top-tier language models, which are limited in the tasks they can perform reliably and in the languages they output. If more actors invest in state-of-the-art generative models, then this could increase the odds that propagandists gain access to them. It is also possible that uncertain and risky investments could lead to the creation of systems that are much better at tasks relevant to influence operations.
    \item The existence of unregulated tooling. Proliferation of easy-to-use interfaces to generate persuasive text or images can increase the adoption of generative models in influence operations. If these tools are developed, we are likely to see an earlier and broader uptick of generated content in influence operations.
    \item Intent-to-use generative models for influence operations. As access to generative models increases, an actor’s willingness to use these models in influence operations might be an important constraint. If social norms do not constrain the use of models to mislead, then actors may be more likely to deploy models for influence operations.
\end{enumerate}

\textit{Willingness to Invest in Generative Models}

In \hyperref[sssec:general]{Section 4.2.2}, we outlined ways that language models could be leveraged for influence operations. First, propagandists could repurpose (or steal) state-of-the-art models with new and more general capabilities. Second, sophisticated propagandists could train models specifically for influence operations. In both cases, the application of generative models to influence operations may ultimately be constrained by different actors’ willingness to make large and potentially risky investments in developing generative models.

To have an impact on influence operations, a large investment need not target generative models for influence operations specifically. An investment could simply target more general generative models for other purposes such as scientific discovery or commercial value. If many actors—such as governments, private firms, and even hyperwealthy individuals—develop these state-of-the-art language models, then that increases the odds that propagandists could gain access (legitimately or via theft) to models that can be repurposed for influence operations. For example, a propagandist could fine-tune a stolen model to produce persuasive text in different languages or in a particular domain.

In the extreme case, the propagandist themself could be a well-resourced actor—like a determined country—and make a risky and large investment in developing a generative model-based system specifically for influence operations. This may require extensive computational resources, bespoke data—such as user engagement metrics—and engineering talent. In either case, it may not be clear how feasible some engineering projects are; the timeline for advances may ultimately depend on whether propagandists decide to make uncertain investments in developing these generative models.

While there are reasons why well-resourced actors might make large investments in developing models for influence, there are also reasons to forgo them. We are already reaching the point where the creation of convincing tweet-sized texts can be automated by machines. However, there could be diminishing returns for influence operations for more advanced capabilities, which would make large investments by propagandists specifically unlikely. For example, if most influence operations rely on a deluge of similarly short bits of content to sway attention-bound humans, there may be few incentives to develop generative models that can generate longer pages of human-like text.

\textit{Greater Accessibility from Unregulated Tooling}

Even with nominal access to models, there will likely be some operational know-how required to use them. For example, applying GPT-3 to propaganda tasks requires fiddling with the exact inputs you give the system. To create a photorealistic image a few years ago, a propagandist would have had to run a model themselves on their own infrastructure. But packaging easy-to-use tools that do these tasks has since lowered the operational know-how required to apply generative models to influence operations. Today, anyone with access to the internet can obtain photorealistic AI-generated images from websites such as thispersondoesnotexist.com. AI-generated profile pictures (images of people) are now commonplace in influence operations\autocite{Bond2022} and have also been used for deceptive commercial purposes.\autocite{Goldstein2022} It is quite possible that had this easy-to-use tooling not been developed, influence operations would not have leveraged AI-generated profile pictures to add plausibility to their campaigns, or may not have done so to the same extent.

An analogous lesson may apply to the use of language models for influence operations as well. If easy-to-use tools for language models proliferate, we may see propaganda campaigns rely on language models (that would otherwise not have). Easy-to-use tools that produce tweet- or paragraph-length text could lower the barrier for existing propagandists who lack machine learning know-how to rely on language models. Easy-to-use tools could also lead to the integration of new capabilities, such as automated chatbots deployed to troll targets determined by a bad actor. At the same time, the creation of easy-to-use language model tools could also lead to the proliferation of propagandists. Firms and private individuals who may once have avoided waging propaganda campaigns could now choose to do so because of declining costs.

\textit{Norms and Intent-to-use}

The intent (or lack thereof) may be an important constraint on the application of generative models to influence operations. In the political science literature, a norm is a “standard of appropriate behavior for actors with a given identity.”\footnote{\autocite{Finnemore1988}. Norms involve two components: a prescription (what to do, or what not to do) and parameters (the situations under which the norm applies). For a description of this literature, see \autocite{Shannon2000}.} Scholars describe three stages for a norm to take hold internationally: norm emergence (a norm is built by norm entrepreneurs, or “people interested in changing social norms”\autocite{Sunstein1996}), a norm cascade (more countries rapidly adopt the norm), and internationalization of the norm (a norm becomes widely accepted and taken for granted.\autocite{Finnemore1988}) Studies show that norms constrain different types of state behavior that would be expected to take place by a cost-benefit analysis. International security scholars have argued that norms have powerfully restrained state behavior—from using nuclear weapons, from more routine use of assassinations, and from widespread use of mercenaries.\footnote{Tannenwald famously argued that non-use of nuclear weapons since the bombing of Hiroshima and Nagasaki cannot be explained by deterrence, but rather is the result of a normative prohibition on the use of nuclear weapons. See: \autocite{Tannenwald1999}. (For evidence that challanges this theory, see \autocite{Dill2022}; \autocite{Percy2007}}

The notion that norms can constrain behavior in different facets of domestic and international life may provide a useful lesson for the use of language models for influence operations. Even if an actor has access to models that can easily be repurposed to create persuasive chatbots, and even if this can be done at minimal cost to them, an actor must still decide to actually build and deploy them. Norms could constrain political actors from using language models for influence operations, and they could encourage developers to inhibit the use of language models for influence operations where possible. 

Creating a norm that it is unacceptable to use language models for influence operations will likely require “norm entrepreneurs” to advocate this position. On the international level, this could be a coalition of states creating an agreement that they will not use language models for propaganda purposes. These states could devise mechanisms to punish those who fail to comply with the norm, or to reward those that join the coalition. On a substate level, machine language researchers or ethicists could also create a coalition to develop norms prohibiting the use of language models for influence operations. In fact, several AI researchers penned an open letter condemning activities like GPT-4chan,\footnote{We discussed this incident in \hyperref[sssec:actors]{Section 4.1.1}. In brief, a researcher fine-tuned a publicly accessible language model on 4chan posts and proceeded to automatically post over 30,000 times in three days.} explicitly citing the lack of community norms around the responsible development and deployment of AI as the reason to speak out.\autocite{CondemningGPT4Chan} Likewise, the marketing and PR industries could develop a norm against providing politicians AI-enabled influence operations as a service.

\newpage
\section{Mitigations}\label{sec:mitigations}
\subsection{A Framework for Evaluating Mitigations}

In this section, we move from describing the threat and attempt to outline a series of possible mitigations that could reduce the dangers of AI-enabled influence operations. Our goal here is to present a range of possible mitigations that various stakeholders could take to reduce the threat of AI-powered influence operations. Importantly, these mitigations are meant to be scoped to language models specifically, and we do not aim to articulate all the mitigations that could be taken to reduce the threat of misinformation generally.\footnote{For one example document that has compiled many strategies and resources for anti-misinformation campaigns, see \autocite{Bianco2020}. See also \autocite{Bontcheva2020}.} Nevertheless, it is important to emphasize that, while generative models could help propagandists produce some types of harmful content, influence operations do not need AI models in order to succeed. As such, mitigations discussed here should be viewed as complements to broader and ongoing counter-influence operations efforts.

We group our mitigations based on four “stages” of the influence operation pipeline where they could be targeted: (1) model construction, (2) model access, (3) content dissemination, and (4) belief formation.\footnote{There are other kill chain models that describe the ways disinformation operators conduct campaigns and how this process could be interrupted. See, for instance, \autocite{Sedova2021}; \autocite{Schneier2019}. However, for the purposes of analyzing the impact of AI language models specifically on disinformation, we use this simplified kill chain model.} This grouping reflects that propagandists need four things to successfully use generative language models to shape the information ecosystem: first, there must be AI models capable of generating scalable and realistic-looking text; second, operators must have regular and reliable access to such models; third, operators must have infrastructure in place to disseminate the outputs of those models; and fourth, there must be a target audience that can be influenced by such content.

\begin{figure}
    \centering
    \includegraphics[width=\textwidth,height=\textheight,keepaspectratio]{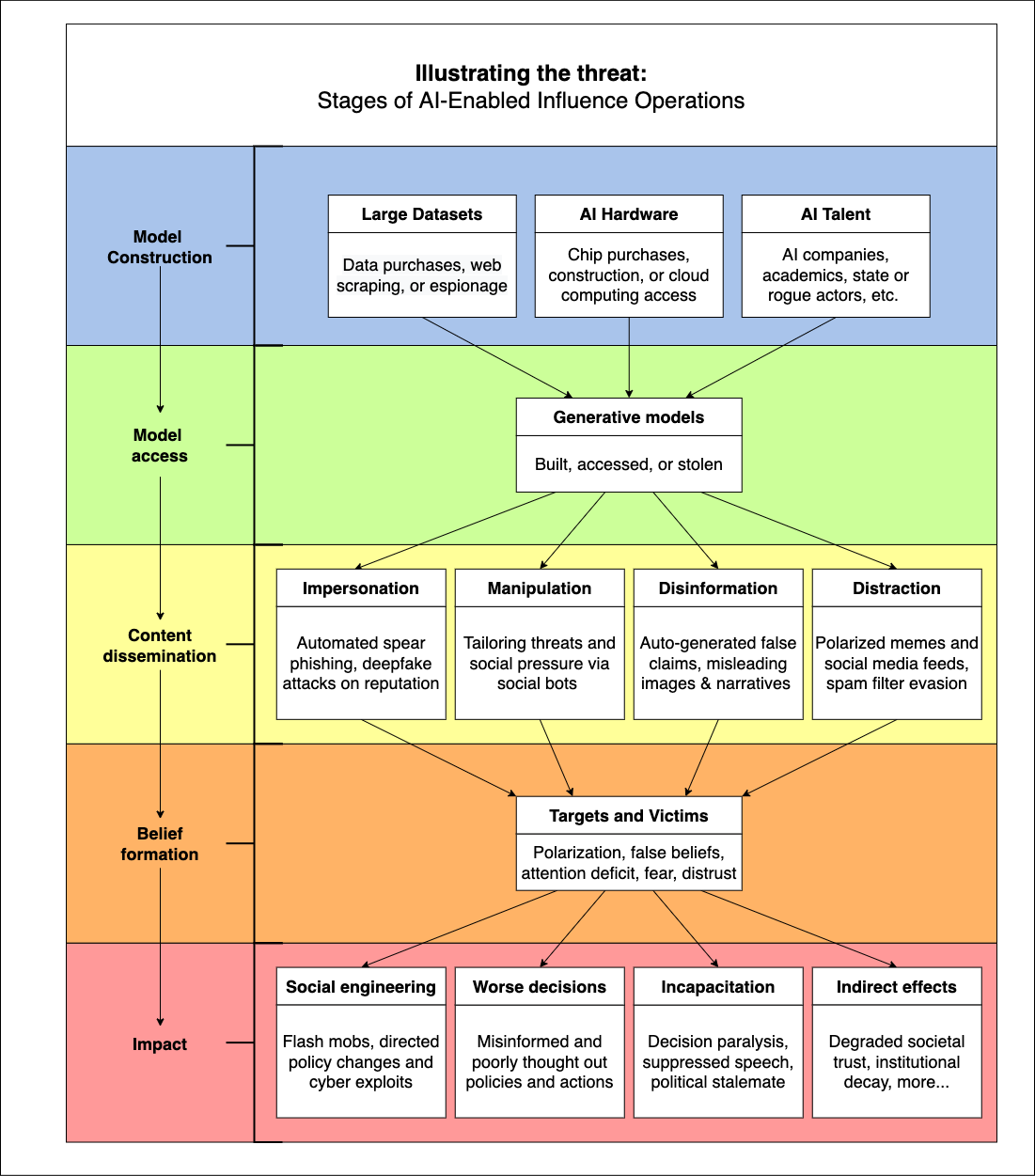}
    \caption{Stages of intervention of AI-enabled influence operations. To disrupt a propagandist’s use of language models for influence operations, mitigations can target four stages: (1) Model Design and Construction, (2) Model Access, (3) Content Dissemination, and (4) Belief Formation. Ultimately, intervening at these stages attempts to mitigate both the direct and indirect effects of influence operations.}
    \label{fig:intervention}
\end{figure}

In Figure \ref{fig:intervention}, we illustrate these points of intervention. For example, a threat actor can use generative model capabilities by accessing a model directly, building it themselves, or stealing the model. Any mitigation that intervenes at the \textbf{Model Access} stage should impact one or more of those three avenues.

For each of these stages, we can think about how an influence operation might be disrupted by using the following sets of questions as starting points:

\begin{itemize}
    \item \textbf{Model Design and Construction:} How could AI models be built so they are robust against being misused to create disinformation? Could governments, civil society, or AI producers limit the proliferation of models capable of generating misinformation?
    \item \textbf{Model Access:} How could AI models become more difficult for bad actors to access for influence operations? What steps could AI providers and governments take?
    \item \textbf{Content Dissemination:} What steps can be taken to deter, monitor, or limit the spread of AI-generated content on social media platforms or news sites? How might the “rules of engagement” on the internet be altered to make the spread of AI-generated disinformation more difficult?
    \item \textbf{Belief Formation:} If internet users are ultimately exposed to AI-generated content, what steps can be taken to limit the extent to which they are influenced?
\end{itemize}

We evaluate each mitigation by paying specific attention to four categories: (1) technical feasibility, (2) social feasibility, (3) downside risk, and (4) impact—four key considerations that stakeholders should use to assess the desirability of pursuing any particular mitigation. In more detail:

\begin{itemize}
    \item \textbf{Technical feasibility} refers to the ability to implement a proposed mitigation on a technical level, without regard to social or political considerations. Some mitigations admit mature and low-cost technical solutions, while others require technical abilities that do not exist, are under question, or would require massive changes to existing technical infrastructure.
    \item \textbf{Social feasibility} refers to the political, legal, and institutional feasibility of a particular mitigation, assuming that it is technically possible to implement. The following questions serve as useful guides for assessing this metric: (1) Can the mitigation be successfully implemented unilaterally, without coordination across multiple independent actors? (2) Do the key actors who could implement the proposed mitigation have incentives in favor of doing so? (3) Would the proposed mitigation be actionable under existing law, regulation, and industry standards? Social feasibility will likely vary by region of interest.
    \item \textbf{Downside risk} refers to the negative impacts, including via negative externalities and second-order effects that a mitigation may cause. Notable downside risks that apply to multiple potential mitigations include heightened forms of censorship, the risk of the mitigation itself being politicized, and the risk of bias (such as inadvertently promoting certain perspectives, cultures, or languages over others).
    \item Finally, \textbf{impact} attempts to evaluate how effective a proposed mitigation would be at reducing the threat of AI-enabled influence operations. For instance, the mitigation “identify all AI-written text on the internet and remove it” is neither technically nor socially feasible, but if it could be implemented, this strategy would completely mitigate the effect of AI-powered influence operations (and thus have high impact). By contrast, “warn people about the dangers of AI-authored content” is much more feasible—but also far less impactful for reducing the effect of AI influence campaigns.
\end{itemize}

Of note, we do not attempt to separate this list of mitigations into “worth trying” and “fine to ignore” categories. Individual stakeholders capable of implementing any of these strategies must weigh the pros and cons of doing so. 
We also encourage additional research to address mitigations that fall outside of our model. We do not lay out mitigations that could shape the distribution of threat actor intentions (e.g., norm development, threats of retaliation) nor that could reduce harms that result from new beliefs shaped by a successful influence campaign. These warrant additional attention, but are not captured by our model.

\begin{table}
\small
\begin{tabular}{| L{0.11\linewidth} | L{0.27\linewidth} | L{0.27\linewidth} | L{0.27\linewidth} |}
    \cline{3-4}
          \multicolumn{1}{c}{} & \multicolumn{1}{c|}{} & Promise; if implemented... & Limitation  \\\hline
         \cellcolor{blue!25} & \cellcolor{blue!25}AI Developers Build Models With More Detectable Outputs & Influence operations with language models will be easily discoverable & Technically challenging and requires coordination across developers \\\cline{2-4}
         \cellcolor{blue!25} & \cellcolor{blue!25}AI Developers Build Models That Are More Fact-Sensitive & Language models will be less effective at spreading falsehoods & Technical methods are still being explored; may only impact some influence operations \\\cline{2-4}
         \multirow{-3}{\linewidth}{\cellcolor{blue!25}Model Design \& Construction } & \cellcolor{blue!25}Developers Spread Radioactive Data to Make Generative Models Detectable & Makes it easier to detect if content is AI generated & Technically uncertain and may be easily circumvented  \\\cline{2-4}
         \cellcolor{blue!25} & \cellcolor{blue!25}Governments Impose Restrictions on Training Data Collection & Limits creation of new models (but only for those in jurisdictions that comply) & Data access restrictions would require high political will  \\\cline{2-4}
         \cellcolor{blue!25} & \cellcolor{blue!25}Governments Impose Access Controls on AI Hardware & Prevents some future models from being developed altogether & Restrictions on semiconductors could escalate geopolitical tensions and hurt legitimate businesses \\\hline
         \cellcolor{green!25} & \cellcolor{green!25}AI Providers Impose Stricter Usage Restrictions on Models & Makes it more difficult for propagandists to obtain cutting-edge models for campaigns & Requires coordination across AI providers and risks hurting legitimate applications \\\cline{2-4}
         \multirow{-3}{\linewidth}{\cellcolor{green!25}Model Access} & \cellcolor{green!25}AI Providers Develop New Norms Around Model Release & Restricts access to future models, but unlikely to prevent propagandists from obtaining already-public ones & Requires coordinating across AI providers and could concentrate capabilities among a small number of companies \\\cline{2-4}
         \cellcolor{green!25} & \cellcolor{green!25}AI Providers Close Security Vulnerabilities & Prevents misuse and access of models via theft and tampering & Only affects one route to model access \\\hline
         \cellcolor{yellow!25} & \cellcolor{yellow!25}Platforms and AI Providers Coordinate to Identify AI Content & Increases the likelihood of detecting AI-enabled influence operations & Will not affect platforms that do not engage; may not work in encrypted channels \\\cline{2-4}
         \cellcolor{yellow!25} & \cellcolor{yellow!25}Platforms Require “Proof of Personhood” to Post & Increases the costs of waging influence operations & Current proof of personhood tests are often gameable by determined operators \\\cline{2-4}
         \multirow{-3}{\linewidth}{\cellcolor{yellow!25}Content Dissemination} & \cellcolor{yellow!25}Entities That Rely on Public Input Take Steps to Reduce Their Exposure to Misleading AI Content & Protects entities relying on public inputs from AI-enabled campaigns & Significant changes to public comment systems could disincentivize participation \\\cline{2-4}
         \cellcolor{yellow!25} & \cellcolor{yellow!25}Digital Provenance Standards Are Widely Adopted & Increases detection of AI-generated content & Significant changes would require large-scale coordination \\\hline
         \cellcolor{red!25} & \cellcolor{red!25}Institutions Engage In Media Literacy Campaigns & Mitigates the impact of influence operations & May reduce trust in legitimate content \\\cline{2-4}
         \multirow{-2}{\linewidth}{\cellcolor{red!25}Belief Formation} & \cellcolor{red!25}Developers Provide Consumer-Focused AI Tools & Increases the likelihood of people consuming high quality information & AI tools may be susceptible to bias; users may become overly reliant on them \\\hline
\end{tabular}
\caption{Summary of Example Mitigations and Selected Promise/Limitation}
\label{tab:mitigations}
\end{table}

In addition, we underscore that we discuss each mitigation in terms of \textbf{who or what institutions} would primarily be responsible for their implementation. But this leaves open the question of \textbf{why} these institutions would implement certain mitigations—specifically, whether they would do so voluntarily or should be compelled by regulators to take certain actions. By framing these mitigations in terms of the enacting institutions, we do not mean to suggest that this problem should be left to the voluntary actions of AI developers and social media platforms. Updated regulations may be called for, and future research could unpack whether government intervention is needed (or desirable) for various mitigations. While we expect mitigations to be applicable across different countries, we focus below specifically on the United States to substantiate our points.

\subsection{Model Design and Construction}

The first stage at which key stakeholders could attempt to disrupt the spread of AI-powered disinformation is when language models are initially conceptualized and trained. How could these models be built differently (or how could they be limited from being built at all) such that it would become harder down the line to use them in influence operations? While the following mitigations might be useful, it is important to emphasize that the ability to construct these models is rapidly proliferating, as discussed in \hyperref[sec:progress]{Section 4}. Since most of these mitigations only affect the development of individual models—and getting consensus on any of these mitigations across all AI developers with the capability of constructing large language models will be very difficult—they generally score low on the metric of social feasibility.

The most reliable method for ensuring that large language models are not used in influence operations is to simply not build large language models. Every other proposed change to the design and construction of these models will be less effective at preventing misuse than not building the model in the first place. However, a complete stop to the development of new large language models is extremely unlikely, and so we focus primarily in this section on how these models could be built \textit{differently} to reduce the risk of misuse.

\subsubsection{AI Developers Build Models With More Detectable Outputs} \label{subsec:detect}

Detecting AI-generated outputs of language models is currently a hard problem that is only getting harder as models improve.\footnote{This is especially true for human detection. For example, researchers found a consistent trend that larger models produce text that is harder to distinguish from human written text; see \autocite{Brown2020}.} However, some actions might be taken based on experiences in other AI subfields to increase the detectability of model outputs. In the subfield of computer vision, researchers at Meta have demonstrated that images produced by AI models can be identified as AI-generated if they are trained on “radioactive data”—that is, images that have been imperceptibly altered to slightly distort the training process. This detection is possible even when as little as 1\% of a model’s training data is radioactive and even when the visual outputs of the model look virtually identical to normal images.\autocite{Sablayrolles2020} It may be possible to build language models that produce more detectable outputs by similarly training them on radioactive data; however, this possibility has not been extensively explored, and the approach may ultimately not work.\footnote{There has been some success demonstrating that radioactive data can be used to induce certain types of behavior in language models; see \autocite{Wallace2020}. However, it is not clear whether radioactive data can be used to generate models whose outputs can be reliably attributed to them.}

Rather than training on radioactive data, statistical perturbations might be introduced to a model’s output by directly manipulating its parameters, thereby distinguishing its outputs from normal text and making detection easier. Past research has identified tools that can be used to detect statistical patterns in outputs from less advanced models such as GPT-2; however, as models become bigger and develop a richer understanding of human text, these detection methods break down if the parameters of the models themselves are not deliberately perturbed in order to enable detection.\footnote{On detection, see \autocite{Gehrmann2019}. However, similar statistical methods perform less well for larger models such as GPT-3 and GROVER; see \autocite{Fröhling2021}. In addition, none of this research assumes a realistic, adversarial threat model, in which attackers are aware that their posts are being assessed to potentially attribute machine authorship. Under this more realistic scenario, attackers could deploy very easy countermeasures, such as altering temperature settings to sample from a wider distribution of possible outputs in order to evade detection.}

However, there are reasons to think that it is difficult to build either highly detectable language models or reliable detection models. Linguistic data—especially across relatively short snippets of text—is already more compressed than in images, with far less room to express the subtle statistical patterns that the Facebook researchers relied on to detect AI-generated images. Still, it is possible that research could identify methods to statistically “fingerprint” a language model.\autocite{Xiang2021} But it is unlikely that individual social media posts will ever be attributable directly to an AI model unless such fingerprints are sufficiently sophisticated: if the patterns permitting such detection were possible, they risk being clear enough for operators to screen out.\footnote{As an example of a trivially circumventable strategy, AI developers could embed special “zero-width” characters in the outputs of their models, which would not immediately be visible to users but which would easily be spotted by automated monitoring tools. There is some research into the use of zero-width characters to attack large language models—see \autocite{Boucher2021, Pajola2021}-but little research into their use as a defensive strategy, in large part because an attacker who was aware that such characters were being inserted into model outputs could easily just remove them before posting content online.} However, these strategies for building more detectable models may still make it possible to attribute larger-scale corpora of text to specific models, though this remains an open question.

Even if some models are designed or redesigned to produce outputs that are traceable at sufficient sizes, attackers could simply gravitate toward other models that are not similarly manipulated. For this mitigation to have a significant impact, it would require high levels of coordination across AI developers who have the ability to deploy large language models. Adversaries with the capability to create their own large language models may merely face additional costs, rather than a loss of capability. Furthermore, operating models that detect whether text is AI-generated represents a challenge, as these will have to be frequently updated to be reliable.

\begin{table}[H]
    \centering
    \begin{tabular}{| >{\raggedright}p{0.2\linewidth} | >{\raggedright\arraybackslash}p{0.7\linewidth} |}
         \hline
         \Centering \textbf{Criteria} & \Centering \textbf{Assessment} \\\hline
         Technical Feasibility & It is an open technical question whether developers will be able to build models that produce detectable outputs. \\\hline
         Social Feasibility & To be implemented effectively, detectable models would require input and coordination across deployers of large language models, which may be socially infeasible. \\\hline
         Downside Risk & There are few obvious downside risks to developing detectable models, assuming there is a low false-positive rate. \\\hline
         Impact & If most or all models are detectable, then influence operations with language models will be easily discoverable. \\\hline
    \end{tabular}
\end{table}

\subsubsection{AI Developers Build Models That Are More Fact-Sensitive}\label{sec:factsensitive}

The dominant paradigm in natural language generation emphasizes “realism” in text generation over other possible values. Models are trained to generate text that effectively mimics (some subsample of) human text, without inherent regard for the truthfulness of the claims that it makes.\footnote{For instance, language models trained on large quantities of internet text will be trained on a large amount of fiction, which can lead them to substitute creative writing for facts.} This means that false claims that are commonly believed may be just as likely for a model to produce as true claims under the current dominant approach to training language models.\footnote{True claims are often a narrow target. Large language models such as GPT-3 are not necessarily truthful by default. See \autocite{Buchanan2021}.}

It may be possible to train AI models in such a way that they are incentivized to make more factually grounded claims, which could produce models that carry less risk of producing falsehoods even if they were accessible to bad actors.\autocite{Evans2021} Significant progress has been made in this area by training models that make use of web searches to improve the factual content of their responses, or that use reinforcement learning techniques to reward more factually correct responses—though these approaches embed their own set of biases about which claims count as “true” or “correct.”\footnote{\autocite{Evans2021}; \autocite{Lowe2022}; \autocite{Hilton2021}.} Other methods attempt to modify the text output to be well-supported by evidence.\autocite{Rashkin2021} While these methods are far from perfect, they can significantly reduce the risk that language models will produce misinformation during ordinary usage.

Nonetheless, most successful influence operations include, or build from, claims that have a kernel of truth.\footnote{As Starbird, Arif, and Wilson write, “To be effective, a disinformation campaign must be based around a ‘rational core’ of plausible, verifiable information or common understanding that can be reshaped with disinformation—for example half-truths, exaggerations, or lies.” See \autocite{Starbird2019}.} Even a language model that produced no false claims could still be used to produce politically slanted or unfalsifiable statements, to shift public attention and discourse, or to engineer false beliefs due to selective context and inauthentic authorship. In fact, in the hands of the right operator, a model that stuck closely to the truth in its outputs might be more persuasive than a model that frequently lied. And further, if this mitigation did meaningfully make it harder for propagandists to misuse language models, it would still require coordination across AI developers to ensure that malicious actors do not simply gravitate toward models that were not trained using similar methods. Finally, to be up to date with the current state of the world, models might have to be retrained very frequently—a requirement that may impose prohibitive costs.

\begin{table}[H]
    \centering
    \begin{tabular}{| >{\raggedright}p{0.2\linewidth} | >{\raggedright\arraybackslash}p{0.7\linewidth} |}
         \hline
         \Centering \textbf{Criteria} & \Centering \textbf{Assessment} \\\hline
         Technical Feasibility & AI developers are exploring ways to make models more fact sensitive, with promising signs of improvement. \\\hline
         Social Feasibility & For the mitigation to be fully implemented, it would require a high degree of coordination between developers of models. \\\hline
         Downside Risk & If language models are more truthful, they may be more persuasive and in turn inadvertently improve the persuasive capabilities of propagandists. \\\hline
         Impact & More truthful language models may be less likely to spread blatant misinformation, but can still serve influence operations relying on true, non-falsifiable, or politically slanted content. \\\hline
    \end{tabular}
\end{table}

\subsubsection{Developers Spread Radioactive Data to Make Generative Models Detectable}

Above, we described that AI developers could attempt to insert “radioactive data” into their datasets when training language models in order to create more detectable outputs. A drawback of this approach is that it requires significant coordination—radioactive data must be inserted by each developer into their own training pipeline. Alternatively, AI researchers, media companies, or governments themselves could choose to proliferate radioactive data directly onto the internet, in locations where it would likely be scooped up by any organization hoping to train a new language model.\footnote{Similar proposals have been advanced in the domain of visual deepfakes, as a way of increasing the likelihood that synthetic images produced from the most common models will be detectable to defenders. \autocite{Hwang2020}.} This would require far less coordination and could potentially make AI outputs more detectable for all future language models. However, this would not affect models that have already been trained, and may be ineffective if developers take steps to filter their training data—a procedure that is common when training models.

This strategy would require proliferators to engage in secretive posting of large amounts of content online, which raises strong ethical concerns regarding the authority of any government or company to deliberately reshape the internet so drastically. In addition, this mitigation would only affect language models trained in the same language in which the radioactive data itself was written. It is also unclear how much of the internet would need to be “radioactive” in this way to meaningfully affect models. And, perhaps most importantly, it remains deeply unclear if this approach would actually result in models with more detectable outputs, for the reasons discussed previously in \hyperref[subsec:detect]{Section 5.2.1}. It seems likely that, even with the use of radioactive training data, detecting synthetic text will remain far more difficult than detecting synthetic image or video content.

\begin{table}[H]
    \centering
    \begin{tabular}{| >{\raggedright}p{0.2\linewidth} | >{\raggedright\arraybackslash}p{0.7\linewidth} |}
         \hline
         \Centering \textbf{Criteria} & \Centering \textbf{Assessment} \\\hline
         Technical Feasibility & While approaches to inserting radioactive data exist for images, it is unclear if this would work for text. \\\hline
         Social Feasibility & A well-resourced actor could unilaterally spread radioactive content that would likely be included in training data for future models. \\\hline
         Downside Risk & Large-scale, secret proliferation of data online raises significant concerns about the desirability of any one group changing the distribution of content on the internet so drastically. \\\hline
         Impact & It is unclear whether this retraining would result in more detectable outputs, and thus detectable influence operations. \\\hline
    \end{tabular}
\end{table}

\subsubsection{Governments Impose Restrictions on Data Collection}

The basis of any large language model is a vast quantity of training data in the form of text generated by real humans. While some of this data is typically taken from relatively structured sources such as Wikipedia, a large majority of data usually comes from tools like Common Crawl that scrape the web for publicly available text.\footnote{CommonCrawl freely publishes its archives of web data. See \autocite{CommonCrawlStart}. But anyone can build their own software for web scraping or use other tools to extract data from websites.} Regulatory or legal changes that would make this type of scraping more difficult to conduct might slow the growth of large language models, while simultaneously forcing developers to focus on extracting information from more structured sources.\footnote{This would in turn have two follow-on effects: learning language from more factually grounded, more formal sources like online news or encyclopedia articles might make models more likely to produce true statements, while also making them significantly less capable of mimicking the language of highly specific target demographics. On using data restrictions to make language models more truthful, see \autocite{Evans2021}: 63.}

%Regulation (EU) 2016/679 of the European Parliament and the Council of 27 April 2016 on the Protection of Natural Persons with Regard to the Processing of Personal Data and on the Free Movement of Such Data, and Repealing Directive 95/46/EC (General Data Protection Regulation), 2016 O.J. L 119/1, art. 14. Major exemptions to this requirement do exist that would likely protect the scraping of textual data for the purposes of scientific research into language models (see ibid., art. 14(5)(b)); however, it is less clear to what extent GDPR may force companies looking to develop commercial AI models to identify impacted data subjects and expressly inform them of their inclusion in a training dataset. Due to the possibility of membership inference attacks on models that could be used to infer personal information about EU citizens, other components of the GDPR relating to protection of personal data may also be implicated in situations where AI developers use web scraping to create training datasets. For research into membership inference, see \autocite{Papernot2016}. (October 2016), https://arxiv.org/abs/1610.05820. At minimum, at least one company has been fined for non-compliance with Article 14 of the GDPR; see \autocite{PolandGDPRFine}. This suggests that even if GDPR does not actually prohibit data scraping (including of personal information) for the purposes of language model construction, companies may feel that it is necessary to spend significantly more on lawyers and compliance efforts to avoid running afoul of the law.}%

These changes could be grounded in changes to federal data privacy laws. Regulations that require internet users to be informed about what their personal data is used for—such as the General Data Protection Regulation (GDPR) in the EU—may slow down large language model development.\footnote{Article 14 of the GDPR requires companies that engage in web scraping of personal information regarding EU citizens to inform data subjects that their personal information has been collected and to grant them certain rights regarding the use of their data. See Regulation (EU) 2016/679 of the European Parliament and the Council of 27 April 2016 on the Protection of Natural Persons with Regard to the Processing of Personal Data and on the Free Movement of Such Data, and Repealing Directive 95/46/EC (General Data Protection Regulation), 2016 O.J. L 119/1, art. 14. Major exemptions to this requirement do exist that would likely protect the scraping of textual data for the purposes of scientific research into language models (see ibid., art. 14(5)(b)); however, it is less clear to what extent GDPR may force companies looking to develop commercial AI models to identify impacted data subjects and expressly inform them of their inclusion in a training dataset. Due to the possibility of membership inference attacks on models that could be used to infer personal information about EU citizens, other components of the GDPR relating to protection of personal data may also be implicated in situations where AI developers use web scraping to create training datasets. For research into membership inference, see \autocite{Papernot2016}; and \autocite{Shokri2016}. At minimum, at least one company has been fined for non-compliance with Article 14 of the GDPR; see \autocite{PolandGDPRFine}. This suggests that even if GDPR does not actually prohibit data scraping (including of personal information) for the purposes of language model construction, companies may feel that it is necessary to spend significantly more on lawyers and compliance efforts to avoid running afoul of the law.} At the extreme end, governments could try to prohibit organizations from mass scraping the web for content at all. More targeted measures could aim at improving cybersecurity for personalized data on social media platforms or prohibiting foreign acquisition of major platforms.\footnote{See \autocite{Helmus2021}; Chapter 1 in \autocite{Schmidt2021}, 50, 405; and \autocite{NationalLawReviewSensitivePersonalData}.}

These mitigations are significantly out of step with the current regulatory environment in the United States, which has not yet passed any comprehensive data privacy laws.\autocite{Klosowski} The Supreme Court has also recently ruled that scraping publicly available data from the web, even in violation of a terms of service agreement, does not violate the Computer Fraud and Abuse Act, the primary cybersecurity law in the United States.\autocite{vanburen2021} Moreover, comprehensive data privacy laws that significantly affect the ability of language model developers to collect data may have large effects in other industries, while also having an uncertain ability to constrain developers outside of the United States. If implemented poorly, data protection measures may harm researchers’ ability to detect and develop countermeasures against influence campaigns more than they hinder campaign planners.\autocite{Bliss2020}

Beyond language models, it may be more feasible to regulate the collection or resale of image or video data. Specific state-level laws, like the Illinois Biometric Information Privacy Act (BIPA), restrict the ability of AI developers to scrape specific types of data—most often pictures of private individuals’ faces—without informed consent.\footnote{Biometric Information Privacy Act, 740 Ill. Comp. Stat. § 14/1–25 (2008).} Such laws have occasionally resulted in successful legal action against AI developers, as when the ACLU successfully used BIPA to compel Clearview AI to screen out data from Illinois residents in its model training pipeline and to sharply limit access to its facial recognition tools within Illinois.\footnote{ACLU v. Clearview AI, Inc., 2020 CH 04353 (Cir. Ct. Cook City., Ill.).} Limiting access to relevant training data can meaningfully disrupt the creation of models that can later be used maliciously; at the same time, to the extent that such limitations are possible at all, they will likely be feasible only for certain restricted sets of training data, such as social media posts or images of private individuals’ faces.

\begin{table}[H]
    \centering
    \begin{tabular}{| >{\raggedright}p{0.2\linewidth} | >{\raggedright\arraybackslash}p{0.7\linewidth} |}
         \hline
         \Centering \textbf{Criteria} & \Centering \textbf{Assessment} \\\hline
         Technical Feasibility & Governmental policy to penalize data collection is likely possible without technical innovation; however, preventing access to internet-based training data is likely difficult. \\\hline
         Social Feasibility & More extreme forms of data access restrictions would require high political will. \\\hline
         Downside Risk & Limiting training data will negatively harm legitimate industries that may rely on language models or their training data and could undermine future detection models. \\\hline
         Impact & Without restricting data collection for all actors, impact is likely limited. \\\hline
    \end{tabular}
\end{table}

\subsubsection{Governments Impose Controls on AI Hardware}

Another path toward limiting the construction of large language models involves either limiting access to or monitoring the usage of AI hardware.\footnote{See, for example, \autocite{Brundage2020}} This could be achieved in a number of ways, including restrictions on the amount of computing power that individual organizations can use to train AI models, disclosure requirements for all AI projects requiring more than a certain threshold of computing power, or export controls on specialized chips.

Monitoring computing power usage may be difficult; some estimates suggest that a model 200 times larger than the current largest language model could be trained using less than 0.5\% of worldwide cloud computing resources.\autocite{Lohn2022} Even if major expenditures of computing power could reliably be identified and tracked, this power is a highly general resource; there is currently little way to tell that an organization purchasing a large amount of computing power is planning to train a large language model as opposed to, say, running climate simulations. However, increasing differentiation between AI compute and non-AI compute could make this easier in the future.\footnote{As one example, AI training may use lower-precision chips; see \autocite{Narasimhan2022}}

Monitoring for large models is currently a difficult task, but semiconductor manufacturing equipment (SME) export controls or restrictions on access to cloud computing resources are easier to implement. In October 2022, the US government announced export controls on semiconductors, SMEs, and chip design software directed at China.\autocite{USDOC2022China, USDOC2022ExportControls} These controls could slow the growth in computing power in China, which may meaningfully affect their ability to produce future language models. Extending such controls to other jurisdictions seems feasible as the semiconductor supply chain is extremely concentrated.\autocite{Khan2020} Another (not mutually exclusive) restriction could involve mandating (or cloud computing companies could voluntarily implement) approval processes for projects requiring enough computing power to build a sophisticated language model. Even simply mandating stock and flow accounting of high-end AI chips could help identify which actors are capable of producing large language models.

To be effective, export controls on computing hardware need to be properly enforced and handle cases such as stockpiling of chips, re-exports via other jurisdictions, and so on. Computing hardware restrictions could also incentivize nation-states to accelerate their indigenous production of AI chips, though some reports argue that it is infeasible for China to scale up the domestic production of SME.\autocite{Khan2020} Furthermore, for the purpose of controlling language model development (or even AI development), export controls on hardware are a blunt instrument and have far-reaching consequences on global trade and many non-AI industries.\autocite{Scheineider2022} Finally, it is worth keeping in mind that often the most impactful propagandists—governments themselves—are those with the capability to plausibly circumvent the hardware restrictions mentioned above.

\begin{table}[H]
    \centering
    \begin{tabular}{| >{\raggedright}p{0.2\linewidth} | >{\raggedright\arraybackslash}p{0.7\linewidth} |}
         \hline
         \Centering \textbf{Criteria} & \Centering \textbf{Assessment} \\\hline
         Technical Feasibility & Some hardware-related controls would not require any technical innovation; however, this likely varies significantly. \\\hline
         Social Feasibility & Restrictions on semiconductors and SMEs have been applied to China; cloud computing restrictions could also be done unilaterally or voluntarily. \\\hline
         Downside Risk & Export controls on semiconductors or semiconductor manufacturing equipment could escalate geopolitical tensions and hurt legitimate businesses. \\\hline
         Impact & US export controls would largely affect the development of future language models in other jurisdictions. \\\hline
    \end{tabular}
\end{table}

\subsection{Model Access}

Once models are built, developers can choose how users interact with them. AI providers have some actions available to them that might reduce bad actors’ access to generative language models. At the same time, these actions could be highly costly for organizations looking to commercialize their models and would require large amounts of cooperation across all relevant AI providers to ensure that propagandists could not simply gravitate toward other equally capable models without similar restrictions in place.

\subsubsection{AI Providers Impose Stricter Controls on Language Models}

As discussed in \hyperref[sec:orienting]{Section 2}, the access regimes governing today’s large language models generally fall into one of three categories: fully private, fully public, or private but accessible under restricted conditions, such as the use of gated API access. Access to many of the most powerful current large language models is partially available through APIs, which provides developers with a number of choices regarding potential access or use restrictions that could be imposed upon their models:

\begin{enumerate}
    \item Developers could require potential users to submit the proposed purposes for which they intend to use a model, and revoke access if actual usage appears to diverge too far from this proposal. This type of restriction was originally a core component of OpenAI’s API access regime, though it has since been replaced with a faster, more automated sign-up process.\autocite{Walsh2021}
    \item Even if the above proposal is adopted, individuals granted API access may often seek to build applications—for instance, chatbots—that give other end users the ability to indirectly input text to a model. These types of applications may indirectly expose the model to bad actors. Developers could therefore impose access restrictions that forbid API users from creating applications that give other users the ability to input arbitrary text to the model.
    \item Developers might choose to restrict model access to only trusted institutions, such as known companies and research organizations, and not to individuals or governments likely to use their access to spread disinformation. Huawei initially appears to have intended an access regime along these lines for its PanGu-\textalpha \hspace{.5pt} model.\autocite{Wiggers2021}
    \item Developers could further limit the number of outputs that individual users can generate within a certain period of time, or they could require review of users who seem to be submitting anomalously large numbers of queries. This would limit the scale of influence operations that rely on language models, but might not prevent their use in more tailored cases (such as generating a smaller number of news articles).
    \item Where API access is granted, developers might also impose restrictions on the types of inputs that users are allowed to submit. For instance, the image-generating model DALL•E 2 attempts to screen out user-submitted queries that are intended to produce “violent, adult, or political” outputs.\autocite{OpenAICurbingMisuse} Such efforts may require significant effort to keep them up to date as new controversial issues arise.
\end{enumerate}

This does not represent an exhaustive list of potential access restrictions. All such restrictions, however, share certain downsides. First, effective restrictions may be difficult for developers to implement, especially if they require manual review or appeal processes. Second, organizations looking to commercialize their models have strong incentives to forego onerous review processes on potential customers. Third, user restrictions are only effective if enough institutions implement strong enough access restrictions to box out bad actors; otherwise, propagandists can simply gravitate toward models with less severe restrictions.

In other words, this proposed mitigation has the makings of a classic collective action problem: the most effective outcome requires coordination across multiple actors, each of whom has incentives to default. In addition, the proposal can only be effective so long as there are no publicly released models that are as effective and easy to use as those maintained by AI developers behind API restrictions. However, if public models are sufficient for propagandists, then this mitigation will likely be less effective.

Despite these limitations, strong industry norms—including norms enforced by industry standards or government regulation—could still make widespread adoption of strong access restrictions possible. As long as there is a significant gap between the most capable open-source model and the most capable API-controlled model, the imposition of monitoring controls can deny hostile actors some financial benefit.\footnote{For a quantitative justification as to why, even if there are good public models available, restrictions on access to (better) private models can still impose non-negligible costs on propagandists, see \autocite{MusserWP}.} Cohere, OpenAI, and AI21 have already collaborated to begin articulating norms around access to large language models, but it remains too early to tell how widely adopted, durable, and forceful these guidelines will prove to be.\autocite{CohereBestPracticesDeployingLanguageModels}

Finally, there may be alternatives to APIs as a method for AI developers to provide restricted access. For example, some work has proposed imposing controls on who can use models by only allowing them to work on specialized hardware—a method that may help with both access control and attribution.\autocite{ChenH2019} Another strand of work is around the design of licenses for model use.\autocite{ResponsibleAILicences} Further exploration of how to provide restricted access is likely valuable.

\begin{table}[H]
    \centering
    \begin{tabular}{| >{\raggedright}p{0.2\linewidth} | >{\raggedright\arraybackslash}p{0.7\linewidth} |}
         \hline
         \Centering \textbf{Criteria} & \Centering \textbf{Assessment} \\\hline
         Technical Feasibility & Some AI developers already restrict usage of models behind APIs. \\\hline
         Social Feasibility & Limiting how AI providers’ language models are used reflects a collective action problem: it requires coordination across AI providers, each of whom has an incentive to defect. \\\hline
         Downside Risk & Limiting access concentrates more power in the hands of a few AI providers and risks undermining those who could benefit from model use. \\\hline
         Impact & If AI developers are governed by norms of restricted use, it could mitigate the potential of AI-enabled influence operations. However, this assumes comparable open-source model developers do not exist. \\\hline
    \end{tabular}
\end{table}

\subsubsection{AI Providers Develop New Norms Around Model Release}

Traditionally, AI researchers have felt bound by what Thomas Merton referred to as the “communism of the scientific ethos,” a norm that holds that a willingness to share information in the interests of full and open collaboration is integral to the scientific enterprise.\autocite{_sectionMerton1973} This norm is not merely a behavioral quirk of scientists; the free and open flow of information is critical for the advancement of science and technology as a whole, and progress in AI has long rested on strong norms of openness and collaboration. But as AI models become increasingly lucrative, this norm is challenged by a competing instinct to privatize models and data in order to commercialize them. In addition, norms of openness in AI research are challenged by safety concerns associated with powerful models that open up new attacks, including the scalable epistemic attacks made possible by powerful language models.\autocite{Liang2022}

Norms regarding data sharing and model release are currently in flux, largely due to progress in large language models. OpenAI has twice broken previous norms regarding model release, first by choosing to delay a full release of GPT-2 in order “to give people time to assess the properties of these models, discuss their societal implications, and evaluate the impacts of release after each stage,” and then again a year later by choosing not to release GPT-3 at all, instead commercializing it behind an API paywall.\autocite{Radford2019} Both of these decisions drew serious criticism at the time, though the use of an API in lieu of a full model release now appears to be somewhat common among AI providers capable of producing cutting-edge language models.\footnote{\autocite{Howard2019}; \autocite{Lipton2019}; \autocite{Dixon2021}.} In the domain of text-to-image models, a sitting member of Congress recently urged the US National Security Advisor and the acting director of the Office of Science and Technology Policy to address the “unsafe releases” of text-to-image models that do not have content restrictions, because they have been used to generate dangerous images.\autocite{EshooToSullivanAndNelson}

While we do not make specific claims about the substance of desirable research norms, a growing body of research is dedicated to examining the types of norms that could be developed to govern AI research, especially in the sphere of large language models. These norms could include staged release of models, the adoption of tradeoff frameworks to assess the risks of open-sourcing models, mechanisms for accepting public feedback and reports of misuse, and prepublication safety review.\autocite{Solaiman2019, Ovadya2019} Implementing any of these norms may require new institutional mechanisms, such as a Partnership on AI-style\autocite{PartnershipOnAI} organization for natural language processing researchers, the creation of a clear set of principles around issues like data collection and model release for large language models, or formal principles regarding what type of risk assessment is expected of AI developers prior to model release.\footnote{For one example of risk assessment for synthetic media, see \autocite{C2PASpecificationsHarmsModelling}.} These institutional mechanisms could help solidify new norms around model design, model release, and model access and would have the potential to significantly impact the ability of propagandists to make use of large language models.

\begin{table}[H]
    \centering
    \begin{tabular}{| >{\raggedright}p{0.2\linewidth} | >{\raggedright\arraybackslash}p{0.7\linewidth} |}
         \hline
         \Centering \textbf{Criteria} & \Centering \textbf{Assessment} \\\hline
         Technical Feasibility & This mitigation does not require technical innovation. \\\hline
         Social Feasibility & The development of norms around language model release for cutting-edge models requires coordination, and open-source developers may choose to ignore those norms. \\\hline
         Downside Risk & Norms that restrict model release may concentrate know-how in the hands of a smaller number of AI providers and impede beneficial AI progress. \\\hline
         Impact & The mitigation would be useful for restricting access to current and future cutting-edge models, but this is unlikely to prevent propagandists from gaining access to already-public models. \\\hline
    \end{tabular}
\end{table}

\subsubsection{AI Providers Close Security Vulnerabilities}

Actors seeking to make use of AI-generated content for propaganda may not be constrained by formal access restrictions to relevant models and research. They may employ covert espionage to steal models and information that will enable construction of their own models, or they may aim to engage in cyberattacks or other forms of sabotage that allow them to manipulate the outputs of already existing language models.\footnote{For a taxonomy of the progression of machine learning vulnerabilities to adversarial influence and a series of case studies on these threats, see \autocite{MitreATLAS}.} For instance, language model poisoning or supply chain attacks on AI providers may allow adversaries to output propaganda from language models they do not possess—manipulating them from afar.\footnote{For instance, in a “model spinning” attack, a threat actor can modify the model to output manipulated narratives whenever a user inputs an adversary-selected trigger word, all without compromising performance. See \autocite{Bagdasaryan2021}. For a general overview of the types of attacks that can be used to target the mathematical peculiarities of AI systems, see \autocite{Lohn2020}.} Similarly, threat actors may also seek to obtain access to cutting-edge, non-public generative models through human vulnerabilities and insider threats at AI institutions.

By developing or hiring groups to simulate adversary attempts to gain access to cutting-edge model capabilities, AI providers can identify and reduce vulnerabilities. Such red-teaming exercises should search not just for cybersecurity vulnerabilities, but also ways in which insider threats or mathematically sophisticated attacks on the AI training process could result in compromised models. Such red teaming can inform a holistic assessment on the risk of the model being misused or applied to produce propaganda. However, while red teaming may successfully identify some vulnerabilities, it is unlikely that all can be caught, and for many types of vulnerabilities that appear to be inherent in modern AI systems, it is unclear how successful any form of technical mitigation can be. Moreover, closing security vulnerabilities is only useful in the context of AI models that have not been made publicly available, as propagandists looking to make use of public models would not need to surreptitiously steal or compromise such models.

\begin{table}[H]
    \centering
    \begin{tabular}{| >{\raggedright}p{0.2\linewidth} | >{\raggedright\arraybackslash}p{0.7\linewidth} |}
         \hline
         \Centering \textbf{Criteria} & \Centering \textbf{Assessment} \\\hline
         Technical Feasibility & Some red-teaming exercises can be performed today, but some defense methods for protecting valuable cyber assets remain research problems. \\\hline
         Social Feasibility & Individual AI developers can implement this mitigation unilaterally. \\\hline
         Downside Risk & There are no obvious downside risks. \\\hline
         Impact & Closing security vulnerabilities is useful insofar as future models are superior for propaganda purposes than already-public models. \\\hline
    \end{tabular}
\end{table}

\subsection{Content Dissemination}

AI-generated content is ultimately only a threat if it reaches and influences real human beings. In general, the interventions most likely to slow the spread of AI-generated propaganda may be those that could be successful against all propaganda, AI-generated or not. However, in this section, we briefly outline a few mitigations that might specifically manage to slow the spread of AI-authored content.

\subsubsection{Platforms and AI Providers Coordinate to Identify AI Content}

It is not clear how companies should respond if or when they judge that content on their platforms was generated by a language model. There are a wide number of plausibly legitimate use cases for AI-generated content on social media, including brand chatbots designed to provide customer service, comedy bots meant to mimic or parody specific authors, or auto-generated news announcements.\footnote{Some of these types of uses already exist; for instance, the account \@dril\_gpt2 on Twitter (https://twitter.com/dril\_gpt2) uses GPT-2 to generate tweets in the style of the dadaist Twitter comedian \@dril.} For this reason, it is unlikely that social media platforms would choose to simply issue a blanket ban on AI-generated content.\footnote{Some social media companies have restrictive policies around the posting of AI-generated images, but even these policies are usually only applicable in certain cases—most commonly, when there is an (assumed) intent to deceive behind the production of the image. See, for instance, \autocite{Bickert2020}, which contains the following explicit exemption: “This policy does not extend to content that is parody or satire, or video that has been edited solely to omit or change the order of words.” The same type of reasons that have led social media companies to avoid adopting blanket bans on AI-generated visual content will also make blanket bans on AI-generated text content unlikely.}

Even if platforms do not issue blanket bans, they might still build in rules regarding appropriate uses of language models into their terms of service. Should accounts generating automated content be required to publicly disclose the origin of content they post? Should posts determined to have been authored by an AI be flagged?\footnote{The impact of flagging content as AI-generated on audiences’ belief formation processes is unknown and may be unintuitive; in one study, for instance, researchers found that survey respondents were just as likely to view “AI-generated” profiles of Airbnb hosts as trustworthy, compared to human-authored profiles. However, when respondents were told that some profiles were human-authored and some were AI-generated, they viewed the profiles they believed were AI-generated as less trustworthy than human-authored profiles. \autocite{Jakesch2019}.} If platforms know that certain external sites host AI-generated content—especially content of a political nature—without disclosing it as such, might that be in itself sufficient grounds to block links to those sites? All of these interventions could be plausible ways to reduce the spread of AI-generated misinformation—assuming it can be identified as such.

Actually detecting content that comes from an AI model, however, is not trivial. Without the aid of AI developers, social media platforms trying to identify machine authorship would be restricted to merely looking for statistical patterns in text and user metadata.\footnote{Humans and machine learning-based detection systems differ in their respective competencies, and can currently perform better at detection together by covering each other’s blindspots. See \autocite{Ippolito2019}.} Current tools for this do not provide the level of confidence that would likely be required for platforms to take disruptive action against accounts, do not work on texts the length of a typical social media post, and are likely to perform worse as models improve.\footnote{One possible statistical method for identifying AI-generated text is provided by \autocite{StrobeltCatchingUnicorns}. But this method assumes that language models will sample text from a relatively constrained distribution, such that the likelihood of unpredictable word patterns ends up significantly lower than is observed in authentic human text. As language models become larger, however, they become capable of accurately modeling a larger distribution of text, decreasing the risk that they will fall into noticeable “most-likely-next-word” ruts. Additionally, many language models permit users to directly manipulate a “temperature” setting, which directly serves to sample from a more unpredictable range of next word outputs when generating text, thereby evading this detection tool more directly.}

However, collaboration between platforms and AI companies may make detection of larger-scale campaigns using AI generation more feasible. For instance, model owners might store outputs so that they can be traced back to the users who generated them.\footnote{This strategy will necessarily be imperfect, as propagandists can always make small or trivial changes to model outputs before posting them to social media. If detection relies on hash matches, operators may easily evade detection by doing so. However, not all operators may be savvy enough to realize that detection makes use of hashes, so this strategy may still have some usefulness. Relying on close-but-not-exact matches to output text, by contrast, introduces a higher level of statistical uncertainty in attribution, though at sufficient scales, campaigns with slightly altered text could still be linked to the use of an AI model with meaningful confidence.} Social media companies could then flag content on their platforms that they suspect may be inauthentic and work with AI companies to determine if any was generated by a language model. This type of collaboration could have follow-on benefits: once an AI company ascertains that a user is reposting outputs to social media, they can work with platforms to determine if other content generated by that user has been reposted to other social media platforms, potentially catching other coordinated inauthentic accounts that the platforms may initially have missed.

This strategy would miss content that is posted to encrypted social media platforms, such as WhatsApp channels. In addition, disinformation is also posted to social media platforms that do not support robust search features and are unlikely to cooperate with AI companies to monitor content, such as Parler and Gab, though it may still be possible to scan public posts on these sites for potential AI-generated content.\footnote{For analyses of Parler and Gab, including an overview of the extent of their content moderation practices, see \autocite{Thiel2021} and \autocite{Thiel2022}.} Without collaboration from the platforms themselves, this mitigation strategy may have only a limited impact.

Despite these drawbacks, partnerships between platforms and AI companies have certain advantages. Unlike imposing onerous up-front access restrictions, this type of monitoring is less likely to alienate potential users from signing up for API access to a language model, which may make it more attractive to AI companies. While bad actors may want to avoid using AI services that engage in this type of monitoring, AI companies can more easily maintain some secrecy about how they monitor for reposted content, making it harder to evade monitoring mechanisms.

\begin{table}[H]
    \centering
    \begin{tabular}{| >{\raggedright}p{0.2\linewidth} | >{\raggedright\arraybackslash}p{0.7\linewidth} |}
         \hline
         \Centering \textbf{Criteria} & \Centering \textbf{Assessment} \\\hline
         Technical Feasibility & Versions of this implementation may be feasible for monitoring publicly posted content, but may be infeasible for encrypted social media channels. \\\hline
         Social Feasibility & Coordination between AI developers and social media companies requires a significant number of bilateral partnerships. \\\hline
         Downside Risk & There are few obvious downside risks assuming the detection models are accurate. If not, they risk flagging the wrong accounts. \\\hline
         Impact & The impact depends on the extent of collaboration between platforms and AI companies, and will not cover all social platforms. \\\hline
    \end{tabular}
\end{table}

\subsubsection{Platforms Require “Proof of Personhood” to Post}

Current policies regarding social media usage range from not requiring any form of registration to requiring that accounts be affiliated with real names and unique email addresses, and, at times, requiring users to submit “video selfies” for proof of personhood.\autocite{InstagramVideoConfirmIdentity} However, any of these approaches can be circumvented by malicious actors: they can register many “burner” email addresses to create fake accounts and hire inexpensive labor to complete proof of humanness checks. 

Platforms could, however, more uniformly require higher standards of proof of personhood in order to verify that content is not being produced by an AI and reposted to their sites. This could involve requiring more reliable forms of authentication when users sign up for an account, for instance, by asking a user to take a live video of themselves posing, or asking for some alternative form of biometrics. Alternatively, platforms could require users to occasionally pass tests to demonstrate humanness before posting content; these tests could either be administered randomly, at periodic intervals, or when a particular user is posting at a high volume. CAPTCHAs are one way to demonstrate humanness in this way; however, a determined adversary can cheaply circumvent them. Outside of tests, another proposed approach includes decentralized attestation of humanness.\autocite{InternetOfHumans}

This mitigation would not make it impossible for propagandists to copy-paste content from a language model into a social media platform and post it. Instead, it would be meant to disrupt operational setups that rely on bots that directly query and post content from language models without explicit human intervention. While this may only describe a minority of influence operations, having such a fully automated capability might be useful to propagandists; for instance, an account could be configured to query a language model every few hours or days for an anodyne post with the intention of posting it directly to a social media platform. Operators would then need only log in every so often to post more explicitly political content, having fully automated the problem of enmeshing those political posts in a more realistic-seeming environment of unrelated content. Requiring checks to post content could meaningfully disrupt this type of operational setup.

There are several significant limitations to this mitigation, including potential infringements on privacy, limits to the types of operations it would mitigate, and limits to its effectiveness against operations by determined adversaries. First, from a privacy perspective, user authentication requirements would likely face resistance from users who are accustomed to an expectation of anonymity online, including users who hold such expectations for very legitimate reasons. Second, hummanness verifications are designed to address operations that rely on social media accounts to spread generated content, but do not affect other information channels—like email or fake news websites. Third, as mentioned above, for well-resourced actors like the Internet Research Agency, the costs of proof of humanness requirements may not be meaningful deterrents: purchasing a new SIM card or hiring cheap outsourced labor to pass a video test will not prevent these campaigns.

Finally, this mitigation introduces an underexplored potential for backlash: If platforms include a proof of humanness check, and propagandists pass such a check, the successful completion could increase the perceived credibility of the account—increasing the persuasive effect from the account in question. Future research could address this question directly.

\begin{table}[H]
    \centering
    \begin{tabular}{| >{\raggedright}p{0.2\linewidth} | >{\raggedright\arraybackslash}p{0.7\linewidth} |}
         \hline
         \Centering \textbf{Criteria} & \Centering \textbf{Assessment} \\\hline
         Technical Feasibility & Various forms of human authentication have been piloted (and implemented) already. \\\hline
         Social Feasibility & Social media platforms and other websites can implement this mitigation unilaterally. \\\hline
         Downside Risk & More extreme forms of this mitigation would undermine online anonymity, which can stifle speech and undermine other human rights. \\\hline
         Impact & The impact depends on the specific implementation: basic CAPTCHA-like tests are gameable, but more novel implementations may increase costs of waging AI-enabled influence campaigns. \\\hline
    \end{tabular}
\end{table}

\subsubsection{Entities That Rely on Public Input Take Steps to Reduce Their Exposure to Misleading AI Content}

Many entities in society rely on public input for feedback, evidence of group beliefs, and legitimacy. For example, when making decisions that affect the community, local planning commissions often seek public comment to make informed decisions.\footnote{In the US context, each branch of the US government has mechanisms for soliciting input from members of the public. For Congress, the most common form of input is constituent calls or emails to their representatives; for the judicial system, the amicus brief provides a means for non-parties to a case to comment on its merits; and for executive agencies, the period of public comment required by the Administrative Procedures Act (APA) allows agencies to understand how affected parties might view proposed regulations.} Similarly, private firms often ask for feedback on products, and media outlets often ask for tips on the issues of the day. The processes that these entities use for public comment constitute potential vectors for the abuse of language models to generate “comments” from the public in order to sway policymakers, local officials, or private entities.

Indeed, there have already been cases in which mass inauthentic comment campaigns have been identified in the US government, most notably when various technology companies submitted millions of comments to the FCC in 2017 regarding net neutrality, falsely using real customers’ names to provide a veneer of legitimacy to the comments.\autocite{WiredISPsFundedFakeComments} Comments generated by a large language model would be more difficult to identify as coordinated, since the comments in the FCC case followed a standard output and merely swapped synonyms for one another. As such, some level of reform to mechanisms for soliciting public input may be called for.

At the lowest end, this reform could simply involve making entities that solicit public comment more aware of the potential for inauthentic content being submitted that poses as public opinion. At the same time, this may have negative externalities: priming policymakers to be suspicious of public input, for example, may itself undermine democratic responsiveness.\autocite{George2020} Organizations soliciting public input might instead choose to implement stronger methods than common CAPTCHAs to ensure that public comments are authentic; currently, many US agencies simply assume that comments are legitimate and perform no follow-up on submitted comments.\footnote{\autocite{AbusesOfFederalNoticeAndComment}; \autocite{FederalRulemakingPublicDataLimitations}. The GAO study found that, for some agencies, as many as 30\% of individuals whose email addresses were associated with public comments reported not having written the comment submitted under their name. Many other agencies did not require email addresses or other types of identifying information for submitted comments, significantly reducing the ability of the agency to authenticate the identity of the commenter.} Here, entities inviting comment will have to ensure that attempts to prevent AI-generated comments do not create frictions that prevent members of the public from participating.\footnote{In the US context, a stronger version could be that the APA itself is amended to mandate some level of vetting for the authenticity of public comments, or criminal liability could be imposed for institutions found to be impersonating members of the public. We do note, however, that the Administrative Conference of the United States (ACUS) has so far preferred not to propose any sweeping changes to the period for public comment. In part, this is because ACUS believes that AI-generated comments could have valuable use cases in the public comment process, such as by generating summaries of public comments or lowering barriers to submitting public comments. See \autocite{George2020}}

\begin{table}[H]
    \centering
    \begin{tabular}{| >{\raggedright}p{0.2\linewidth} | >{\raggedright\arraybackslash}p{0.7\linewidth} |}
         \hline
         \Centering \textbf{Criteria} & \Centering \textbf{Assessment} \\\hline
         Technical Feasibility & Basic defenses—like user authentication—to prevent bots from overwhelming public comment boards already exist. \\\hline
         Social Feasibility & Policy change will likely require coordination across multiple parts of government. \\\hline
         Downside Risk & Significant changes may disincentivize members of the public from participating in public comment periods. \\\hline
         Impact & The impact varies depending on the specific implementation, but could make public input solicitation much more robust. \\\hline
    \end{tabular}
\end{table}

\subsubsection{Digital Provenance Standards Are Widely Adopted}

Because technical detection of AI-generated text is challenging, an alternate approach is to build trust by exposing consumers to information about how a particular piece of content is created or changed. Tools such as phone cameras or word processing software could build the means for content creators to track and disclose this information.\footnote{For one example of a media provenance pipeline from certified authoring tools to browser extensions for verification, see \autocite{England2021}.} In turn, social media platforms, browsers, and internet protocols could publicize these indicators of authenticity when a user interacts with content.

This intervention requires a substantial change to a whole ecosystem of applications and infrastructure in order to ensure that content retains indicators of authenticity as it travels across the internet. To this end, the Coalition for Content Provenance and Authenticity (C2PA) has brought together software application vendors, hardware manufacturers, provenance providers, content publishers, and social media platforms to propose a technical standard for content provenance that can be implemented across the internet.\autocite{C2PASpecificationsHarmsModelling} This standard would provide information about content to consumers, including its date of creation, authorship, hardware, and details regarding edits, all of which would be validated with cryptographic signatures.\autocite{W3CVerifiableCredentialsDataModel}

Theoretically, this standard would work for AI-generated content, particularly if AI-as-a-service companies opt in to self-declare authorship for each piece of content and require applications or individuals accessing their services through API to do the same. Over time, users may learn to trust the content that has provenance markers and distrust content that lacks them. However, these protocols cannot authenticate preexisting legacy content. In addition, while these measures can provide greater transparency about the creation, history, and distribution of files—including images and text files generated by word processing applications—they cannot provide a means for authenticating and tracking the spread of \textit{raw text}, which can be copied and pasted from file to file without leaving a record in a specific file’s history. To authenticate text provenance widely would require radical changes to internet protocols. For example, it is possible that the HTTP protocol would have to be modified to embed content provenance information. Since language models output raw text and not files, simply storing provenance information in files is sharply limited in its ability to help track the spread of AI-generated misinformation. More low-level changes may be needed to maximize the impact of this intervention.

If the provenance information for a piece of content contains information about the user, then this intervention would raise privacy risks.\footnote{For more discussion of privacy risks here, see \autocite{WitnessTicksOrDidntHappen}.} This implementation could threaten anonymous speech on the internet. However, if only information to distinguish AI and human-generated content is added, then the privacy risks are lower.

\begin{table}[H]
    \centering
    \begin{tabular}{| >{\raggedright}p{0.2\linewidth} | >{\raggedright\arraybackslash}p{0.7\linewidth} |}
         \hline
         \Centering \textbf{Criteria} & \Centering \textbf{Assessment} \\\hline
         Technical Feasibility & Promising technical paths exist, but the technology has not yet been proven. \\\hline
         Social Feasibility & Some progress has been made in coordinating between interested parties, but robust versions of this mitigation would require massive coordination challenges. \\\hline
         Downside Risk & Adding author information raises privacy risks. \\\hline
         Impact & Radical changes to guarantee content provenance would have high impact, but more feasible options would likely have limited impact. \\\hline
    \end{tabular}
\end{table}

\subsection{Belief Formation}

The preceding mitigations address the supply of AI-generated misinformation. However, as long as target audiences remain susceptible to propaganda that aligns with their beliefs, there will remain an incentive for influence operations generally, as well as incentives more specifically for propagandists to leverage AI to make those operations more effective. In this section, we therefore discuss two interventions that might help address the demand side of the misinformation problem: media literacy campaigns, and the use of AI tools to aid media consumers in interpreting and making informed choices about the information they receive.

\subsubsection{Institutions Engage in Media Literacy Campaigns}

There is some evidence that media literacy campaigns can increase individuals’ ability to discern between real and fake news online.\footnote{\autocite{Roozenbeek2020}; \autocite{Guess2020}; \autocite{Jeong2012}; \autocite{Helmus2020}} Existing media literacy tools that teach people how to “spot” coordinated accounts online, however, sometimes emphasize traits or mistakes that AI tools can avoid making, such as repetitiveness or a lack of “personal” content interspersed with more political content.\footnote{For an existing example of a media literacy tool that teaches users the “telltale” signs of troll accounts, see \autocite{SpotTheTrollGame}.} If current programs become outdated, media literacy will require updating. For example, if language models overcome repetition and lack of “personal” content, literacy campaigns can still combat the goals of the propagandists by teaching people to fact-check content in articles and to distinguish objective information from false, misleading, or slanted content.\footnote{For one example of the effectiveness of these measures, see \autocite{Pennycook2021}.} These campaigns may have less impact, however, on distraction operations that crowd out genuine news.

Unlike many of the other mitigations listed above, the impact of media literacy campaigns is agnostic to human versus computer authorship. These efforts focus on teaching people how to analyze content, not necessarily to spot AI-generated content. Another form of digital literacy campaigns could be to teach people about AI-generated content specifically. If new “telltale” signs can be identified that represent common indicators of AI-powered influence operations, then this mitigation could be beneficial. However, if the most that can be said of AI-powered operations is that they look more authentic than human-operated campaigns, then this strategy may be misplaced. Emphasizing that any account on the internet could be an AI-powered bot may make people more likely to simply dismiss arguments they disagree with as inauthentic and not worth paying attention to, thereby exacerbating societal division and polarization. Overemphasizing the prevalence and danger of misinformation online may ultimately serve the same goal that propagandists themselves are often trying to achieve: making people inherently distrustful of any information or argument that conflicts with their preexisting beliefs.\autocite{Hao2019}

\begin{table}[H]
    \centering
    \begin{tabular}{| >{\raggedright}p{0.2\linewidth} | >{\raggedright\arraybackslash}p{0.7\linewidth} |}
         \hline
         \Centering \textbf{Criteria} & \Centering \textbf{Assessment} \\\hline
         Technical Feasibility & No technical innovation is required. \\\hline
         Social Feasibility & A variety of actors could unilaterally lead educational campaigns. \\\hline
         Downside Risk & Educating about the threat of AI-enabled influence operations could reduce trust in genuine content or in online information environments more broadly. \\\hline
         Impact & Educational initiatives could help people distinguish reliable information from misinformation or slanted text, and mitigate the effects of influence operations (AI-generated or not). \\\hline
    \end{tabular}
\end{table}

\subsubsection{Developers Provide Consumer-Focused AI Tools}

Just as generative models can be used to generate propaganda, they may also be used to defend against it. Consumer-focused AI tools could help information consumers identify and critically evaluate content or curate accurate information. These tools may serve as an antidote to influence operations and could reduce the demand for disinformation. While detection methods (discussed in \hyperref[subsec:detect]{Section 5.2.1}) aim to detect whether content is synthetic, consumer-focused tools instead try to equip consumers to make better decisions when evaluating the content they encounter.

Possibilities for such tools are numerous.\footnote{For a variety of examples of consumer-focused tools that help users control the information they see, see \autocite{FTCCombattingOnlineHarms}.} Developers could produce browser extensions and mobile applications that automatically attach warning labels to potential generated content and fake accounts, or that selectively employ ad-blockers to demonetize them. Websites and customizable notification systems could be built or improved with AI-augmented vetting, scoring, and ranking systems to organize, curate, and display user-relevant information while sifting out unverified or generated sources.\footnote{A particularly successful example of a curation tool is Live Universal Awareness Map, which has done near real-time source aggregation on conflicts in Ukraine and Syria while aiming to filter out state-sponsored propaganda. On karma and reputation systems, see \autocite{Seger2020}; and \autocite{Johnson2021}} Tools and built-in search engines that merely help users quickly contextualize the content they consume could help their users evaluate claims, while lowering the risk of identifying true articles as misinformation.\footnote{The issue of false positives—identifying quality sources as misleading or false—is common with social media fact-checking recommendation systems, which often superficially associate new accurate articles with prior false ones, or fail to differentiate between false claims and claims that are contingent, probabilistic, or predictive in nature.} Such “contextualization engines” may be especially helpful in enabling users to analyze a given source and then find both related high-quality sources and areas where relevant data is missing. By reducing the effort required to launch deeper investigations, such tools can help to align web traffic revenue more directly with user goals, as opposed to those of advertisers or influence operators.\autocite{Ovadya2021} Another proposal suggests using  AI-generated content to educate and inoculate a population against misleading beliefs.\autocite{GGF2020HumorOverRumor, Herriman2020}

Some of the most promising AI-enabled countermeasures may leverage state-of-the-art generative models themselves, to reshift the offense-defense balance in favor of information consumers.\footnote{By tailoring to serve the needs of individual information consumers, such tools could equip consumers with decision-informing capabilities that would otherwise be too risky to implement at the scale of an entire platform.} As generative models get better at producing persuasive arguments that exploit viewer biases and blindspots, defensive generative models could be used to help users detect and explain flaws in tailored arguments or to find artifacts in manipulated images.\autocite{Leike2022} Generative models that help users find relevant information can also be trained how to “show their work” by citing sources that support their answers.\autocite{Nakano2021} Such methods could serve as building blocks for future tools that augment a consumer’s ability to critically evaluate information.

Consumer-focused tools may also go beyond the individual, with more expensive, AI-enabled intelligence services that offer tools to businesses, governments, and other organizations that aim to increase their awareness of, and improve their responses to, influence operations. 

Despite their prospective benefits, AI tools will also present risks. They are likely to be susceptible to forms of social bias, just as current models are. Defensive generative models that are aligned with consumer incentives may also exacerbate confirmation bias, as consumers may prefer information that tailors to their preexisting biases. Social media companies may make it difficult or against their policies for externally developed tools to interface with their platforms, both to protect privacy and to sustain user engagement. While social media companies may be in a good position to provide their own defensive AI tools, the divergence between their interests and those of their users would likely exceed that of third-party tool providers. Accordingly, tools created by platforms could also serve to discourage more effective policy action and to justify disabling the use of third-party tools that aren’t as aligned with platform objectives.\footnote{For example, the use of such tools could be used to impress Congress with a platform’s efforts, and to make the argument that users already have plenty of options to seek out or control the information they are exposed to, even if in practice the tools are designed to discourage use.}

More powerful versions of web-searching generative models may also pose new unique risks if their range of action and reinforcable behavior is not carefully constrained. For models that are capable of generating and inputting text queries within other websites to find more relevant results, the incentive to return useful results could reward fraudulent behavior (e.g., editing and returning Wikipedia results if there aren’t good sources\autocite{Nakano2021}). While many such specific imagined threats are highly unlikely, the potential impacts of defensive generative models on search engine traffic and the internet itself should be accounted for.

Overall, consumer-focused AI tools provide a variety of opportunities to head off the impact of influence operations that employ stronger generative models, but they will require high-quality implementation.

\begin{table}[H]
    \centering
    \begin{tabular}{| >{\raggedright}p{0.2\linewidth} | >{\raggedright\arraybackslash}p{0.7\linewidth} |}
         \hline
         \Centering \textbf{Criteria} & \Centering \textbf{Assessment} \\\hline
         Technical Feasibility & Creating AI tools that help people reason or highlight factual inaccuracies is an ongoing research problem, but some promising directions exist. \\\hline
         Social Feasibility & Some progress could be achieved unilaterally by researchers or entrepreneurs, but coordination with social media platforms would be required for broader effect. \\\hline
         Downside Risk & AI tools may be susceptible to bias, and people could become overly reliant on them. \\\hline
         Impact & If implemented well, defensive AI tools could have a big impact in helping consumers form accurate beliefs. \\\hline
    \end{tabular}
\end{table}

\newpage
\section{Conclusions}\label{sec:conclusions}

While each of the mitigations discussed above are important to weigh on their own merits, there are some crosscutting conclusions that we offer to policymakers trying to think through the problem of AI-powered influence operations. Our shared assessments of these mitigations lead to the following main conclusions:

\begin{enumerate}
    \item Language models are likely to significantly impact the future of influence operations.
    \item There are no silver bullets for minimizing the risk of AI-generated disinformation.
    \item New institutions and coordination (like collaboration between AI providers and social media platforms) are needed to collectively respond to the threat of (AI-powered) influence operations.
    \item Mitigations that address the supply of mis- or disinformation without addressing the demand for it are only partial solutions.
    \item More research is needed to fully understand the threat of AI-powered influence operations as well as the feasibility of proposed mitigations.
\end{enumerate}

\subsection{Language Models Will Likely Change Influence Operations}

As outlined in \hyperref[sec:influence]{Section 4}, language models have the potential to significantly affect how influence operations are waged in the future—including the actors waging these campaigns, the behaviors of the propagandists, and the content included.

\begin{adjustwidth}{0.05\linewidth}{0em}
  \textit{Actors:} If generative models become widely accessible, it will drive down the cost of producing propaganda; in turn, those who have refrained from waging influence operations in the past may no longer be disinclined. Private PR and marketing firms may develop knowledge in how to most effectively integrate these models, and thus serve as a resource and scapegoat for political actors seeking to outsource their campaigns.
  
  \textit{Behaviors:} Language models offer to change how influence operations are waged. They may be deployed for dynamic generation of responses, automated cross-platform testing, and other novel techniques. Although we described a few new possible behaviors in this report, we suspect propagandists will use these models in unforeseen ways in response to the defensive measures that evolve.
  
  \textit{Content:} Language models will likely drive down the cost and increase the scale of propaganda generation. As language models continue to improve, they will be able to produce persuasive text—text that is difficult to distinguish from human-generated content—with greater reliability, reducing the need for skilled writers with deep cultural and linguistic knowledge of the target population.
\end{adjustwidth}

Although we foresee these changes in the medium term, there is some speculation at play. The extent to which language models change the nature of influence operations is dependent on critical unknowns, including diffusion and accessibility, and various technical and social uncertainties. We do not yet know who will control these models, and how information environments—like social media platforms—will adapt in a world where models are widely available for use.

\subsection{There Are No Silver Bullet Solutions}

\hyperref[sec:mitigations]{Section 5} discussed a large number of possible strategies for managing the threat of AI-generated influence operations. Unfortunately, no proposed mitigation manages to be simultaneously (1) technically feasible, (2) institutionally tractable, (3) robust against second-order risks, and (4) highly impactful. The fact that large language models are increasingly proliferating—both behind paid APIs and in the form of openly released models—currently makes it all but impossible to ensure that large language models are never used to generate disinformation.

This is not an excuse for defeatism. Even if responding to the threat is difficult, AI developers who have built large language models have a responsibility to take reasonable steps to minimize the harms of those models. By the same token, social media companies have a continuing obligation to take all appropriate steps to fight misinformation, while policymakers must seriously consider how they can help make a difference. But all parties should recognize that any mitigation strategies specifically designed to target AI-generated content will not fully address the endemic challenges.

Even if better policies can be adopted to govern the majority of language models, very few interventions will stop a well-resourced, non-cooperative state from constructing its own alternatives. One option for countries like the United States would be to soften immigration requirements for AI talent, which could concentrate the ability to produce language models in a few countries—though this too will be unlikely to fully stop a sufficiently motivated nation-state from developing high capability systems of their own.

\subsection{Collective Responses Are Needed}

Many of the mitigations discussed above might have a meaningful impact in reducing AI-generated influence campaigns, but only if new forms of collaboration are developed. Strong norms among the AI community—regarding either the release of models or the training methods used to develop them—could make it harder for the most common language models to be induced to generate disinformation. We have also suggested that if detection of AI-generated text will be feasible at all, it will likely require relatively large “batches” of outputted text in order to attribute. Collaboration between social media companies and AI companies may be necessary in order to curate and attribute large batches of potentially inauthentic content.

The current US response to influence operations is fractured: fractured among technology companies, fractured among academic researchers, fractured between multiple government agencies, and fractured on the level of collaboration between these groups. Social media companies have different approaches to whether (and how) to treat influence operations; academics lack relevant data to understand related issues; AI developers often lack sufficient expertise to understand potential abuses of the technologies they create, and responsibilities for influence operations are not clearly delineated to any single US department or agency. Policymakers should consider creating stronger mechanisms and incentives to ensure coordination across all relevant stakeholders.\footnote{The National Security Commission on AI, the Aspen Institute, and a variety of others have recommendations for how to integrate government efforts to counter foreign-sourced influence campaigns. See \autocite{Schmidt2021}; \autocite{Waltzman2017}; \autocite{Schoen2012}; \autocite{Chessen2017}; \autocite{Sedova2021}.}

\subsection{Mitigations Must Address Demand As Well As Supply}

All else being equal, the fact that a particular post was authored by an AI does not in itself make the content of that post less truthful or more destabilizing than the same content would be coming from a human. While this paper has focused on mitigations that would disrupt the pipeline between large language models and influence operations, it is important to emphasize that many other mitigations can be implemented or further strengthened that aim to reduce the spread of false or biased information generally. Some social media platforms have already implemented a number of these mitigations—though often not equitably between English-speaking countries and other regions. But influence operations appear to be a new normal of online activity, and more effort to improve these mitigations is warranted.

It is equally important, however, to emphasize that mitigations that disrupt the supply of misleading information are ultimately only partial solutions if the demand for misleading information remains unchanged. While people rarely demand to be misinformed directly, information consumers often demand information that is cheap and useful for their goals—something influence operations can tailor to with greater freedom from the constraints of reality.

From a selfish perspective, ignorance is often rational: it is not possible to be informed on everything, gathering accurate information can be boring, and countering false beliefs may have social costs.\autocite{Downs1957} Similarly, consuming and sharing disinformation may be entertaining, attract attention, or help an individual gain status within a polarized social group. When the personal costs of effortful analysis exceed the personal benefits, the likely result will be lower-quality contribution to group decision-making (e.g., sharing disinformation, free riding, groupthink, etc.).

\subsection{Further Research Is Necessary}

Many of the properties of large generative models are not fully understood. Similarly, clarity is still missing regarding both the structure and the impacts of many influence operations, which are conducted in secret.

Clarity on the scale of the threat posed by influence operations continues to be elusive. Is the actual impact of such campaigns proportionate to the attention they receive in the popular imagination and press coverage? How effective are existing platform-based mitigations—such as friction measures designed to slow down the virality of content—at reducing the spread of misinformation? As it relates to influence operations with generative models specifically, future research should unpack the differential impact these technologies may have on different populations. For example, relevant factors include the languages various models output most persuasively, and the media and internet fluency in different communities. AI developers and researchers could reach out to communities likely to be impacted to better understand their risks and needs.

A number of technical issues are also currently ambiguous. The relationship between model size, length of fine-tuning, and overall performance or persuasiveness, for instance, is unclear. While it is generally true that larger, more heavily trained models perform better across a wide variety of tasks—including disinformation-related ones—it is not clear whether fine-tuning a smaller model can reliably make up that gap. How do these factors change between models primarily trained on large, well-represented languages like English and those with more capability to use less well-represented languages? On the mitigation side, the feasibility of detection methods remains ambiguous. Although it seems reasonable to assume that (1) attributing short pieces of content as AI-generated will remain impossible and (2) detection might become possible at much larger scales, it is hard to be more specific than this. What scales are necessary to enable detection? How much can perturbing models or training on radioactive data alter this necessary threshold? Furthermore, how realistic is it to train models in ways that reduce their likelihood of outputting misleading content to begin with?

Further research would also be useful to better understand, model, and clarify the decision-making of propagandists themselves. Detailed analyses of the relative gains that malicious actors can capture by incorporating generative models into their operations are also lacking. It is similarly unclear whether API restrictions on large language models meaningfully discourage operators from accessing certain services, and if they do, whether operators are able to simply gravitate toward open-source models without any loss of capability.\footnote{Forthcoming work from some of the authors will attempt to partially address this. See \autocite{MusserWP}}

Finally, this is a rapidly moving field where norms have not yet solidified. Should AI developers release or restrict their models? Should internet researchers publish observed tactics of propagandists or keep them secret? To what extent can platforms and AI developers form meaningful partnerships that can aid in the detection and removal of inauthentic content? At the broadest level, thoughtful engagement with all of these questions—both from people within the relevant industries and from neutral, third-party observers—is a critical necessity.

\newpage
\phantomsection\addcontentsline{toc}{section}{References}
\printbibliography

\end{document}